\documentclass[
twocolumn,
english,
aps,
pra,
longbibliography,
superscriptaddress,
amsmath,
amssymb,
floatfix,
]{revtex4-1}
\usepackage[utf8]{inputenc}
\usepackage{microtype}
\usepackage{graphicx}
\usepackage{amsmath}
\usepackage{amssymb}
\usepackage{mathtools}
\usepackage[colorlinks=true,urlcolor=blue,citecolor=blue,linkcolor=blue]{hyperref}
\usepackage{natbib}
\usepackage{physics}
\usepackage{cleveref}
\usepackage{sidecap}
\usepackage{makecell}
\usepackage{listings}
\usepackage{multirow}
\usepackage{siunitx}
\usepackage[linesnumbered,lined,commentsnumbered]{algorithm2e}

\begin{document}

\title{Quantum Approximate Multi-Objective Optimization}
\date{\today}

\author{Ayse Kotil}
\affiliation{%
IBM Quantum, IBM Research Europe - Zurich
}%
\affiliation{%
Zuse Institute Berlin
}%

\author{Elijah Pelofske}
\affiliation{%
Los Alamos National Laboratory
}%

\author{Stephanie Riedmüller}
\affiliation{%
Zuse Institute Berlin
}%

\author{Daniel J.~Egger}
\affiliation{%
IBM Quantum, IBM Research Europe - Zurich
}%

\author{Stephan Eidenbenz}
\affiliation{%
Los Alamos National Laboratory
}%

\author{Thorsten Koch}
\affiliation{%
Zuse Institute Berlin
}%
\affiliation{%
Technische Universität Berlin
}%

\author{Stefan Woerner}
\email{wor@zurich.ibm.com}
\affiliation{%
IBM Quantum, IBM Research Europe - Zurich
}%

\begin{abstract}
The goal of multi-objective optimization is to understand optimal trade-offs between competing objective functions by finding the Pareto front, i.e., the set of all Pareto optimal solutions, where no objective can be improved without degrading another one.
Multi-objective optimization can be challenging classically, even if the corresponding single-objective optimization problems are efficiently solvable.
Thus, multi-objective optimization represents a compelling problem class to analyze with quantum computers.
In this work, we use low-depth Quantum Approximate Optimization Algorithm to approximate the optimal Pareto front of certain multi-objective weighted maximum cut problems.
We demonstrate its performance on an IBM Quantum computer, as well as with Matrix Product State numerical simulation, and show its potential to outperform classical approaches.
\end{abstract}

\maketitle{}

\section{Introduction\label{sec:introduction}}

Quantum computing is a new computational paradigm that has the potential to disrupt certain disciplines, with (combinatorial) optimization frequently being mentioned as one of them \cite{Abbas2023}. However, in many cases, classical approaches can find good solutions quickly, leaving little room for further improvements. Thus, it is important to identify the right problem classes and instances that are truly difficult classically with a potential for improvement through quantum computers.

The goal of multi-objective optimization (MOO) is to find optimal trade-offs between multiple objective functions by identifying all Pareto optimal solutions, i.e., those solutions where no single objective value can be improved without degrading another one \cite{Ehrgott2005}.
MOO can be difficult even if corresponding single-objective problems can be solved efficiently \cite{Ehrgott2005, Figueira2016}, particularly with an increasing number of objective functions \cite{Allmendinger2022}.
In this context, sampling-based approximate quantum optimization algorithms can be beneficial as they can quickly produce a large variety of good solutions to approximate the Pareto front.

We demonstrate how the quantum approximate optimization algorithm (QAOA) can be efficiently applied to multi-objective combinatorial optimization by leveraging transfer of QAOA parameters across problems of increasing sizes. 
This eliminates the need to train QAOA parameters on a quantum computer and removes a computational bottleneck for the considered problems. 
Using an IBM Quantum computer, we present promising results demonstrating that our algorithm has the potential to outperform classical approaches for multi-objective weighted maximum cut (MO-MAXCUT). Additionally, we show how today's quantum computers can help forecast the performance of heuristics on future devices, serving as tools for algorithmic discovery. 

The remainder of the paper is structured as follows.
Sec.~\ref{sec:moo} introduces MOO. 
Sec.~\ref{sec:classical_algorithms} discusses classical approaches to approximate the Pareto front. 
Then, Sec.~\ref{sec:qaoa} defines the QAOA \cite{farhi2014quantumapproximateoptimizationalgorithm}--a core building block of our quantum algorithm--and the required theory.
Afterwards, Sec.~\ref{sec:quantum_moo} discusses the related work on quantum algorithms for MOO and introduces our approach.
Demonstrations of how the algorithm performs and how it compares to a classical benchmark are presented in Sec.~\ref{sec:experiments}.
Sec.~\ref{sec:conclusion} summarizes our work and discusses possible future research.

\section{Multi-objective Optimization\label{sec:moo}}

The goal of MOO is to maximize a set of $m \in \mathbb{N}$ objective functions $(f_1, \ldots, f_m)$, with $f_i: X \rightarrow \mathbb{R}$, $i \in \{1, \ldots, m\}$, over a feasible domain $X \subset \mathbb{R}^n$, $n \in \mathbb{N}$ \cite{Ehrgott2005}:
\begin{eqnarray}
    \max_{x \in X} && (f_1(x), \ldots, f_m(x)).
\end{eqnarray}
Since it is usually not possible to optimize all $f_i$ simultaneously, MOO refers to identifying the optimal trade-offs between possibly contradicting objectives.

Optimal trade-offs between objectives are defined via the concept of Pareto optimality. A solution $x \in X$ is said to dominate another solution $y \in X$ if $f_i(x) \geq f_i(y)$ for all $i \in \{1, \ldots, m\}$ and if there exists at least one $j \in \{1, \ldots, m\}$ with $f_j(x) > f_j(y)$. A solution $x$ is called Pareto optimal if there does not exist any $y$ that dominates $x$. In other words, $x$ is Pareto optimal if we cannot increase any $f_i$ without decreasing at least one $f_j$, $j \neq i$. The set of all Pareto optimal solutions is called the Pareto front or efficient frontier. Solving MOO exactly denotes computing the full Pareto front. 

Suppose $X$ is a convex set and all $f_i$ are concave functions. Then, the Pareto front usually consists of infinitely many points that lie on the surface of a convex set \cite{Ehrgott2005}. However, for non-convex problems, e.g., in the case of discrete optimization with $X \subset \mathbb{Z}^n$, this is not true anymore, and the Pareto front may consists of a finite discrete set of points \cite{Ehrgott2005}.
We call Pareto optimal solutions supported if they lie on the convex hull of the Pareto front, and non-supported otherwise.
Especially discrete MOO problems are often not efficiently solvable even when the corresponding single-objective optimization problems are in \textbf{P}, i.e., if they can be solved efficiently. This is the case, if the Pareto front grows exponentially with the input size or if the computation of non-supported solutions is demanding \cite{Ehrgott2005, Figueira2016}.
Throughout this manuscript, we usually assume $X = \{0, 1\}^n$ and $f_i(x) = x^T Q_i x$ for matrices $Q_i \in \mathbb{R}^{n \times n}$, i.e., we consider the MOO variant of quadratic unconstrained binary optimization (QUBO), although many of the introduced concepts are more general.

\begin{figure}
    \centering
    \includegraphics[width=\linewidth]{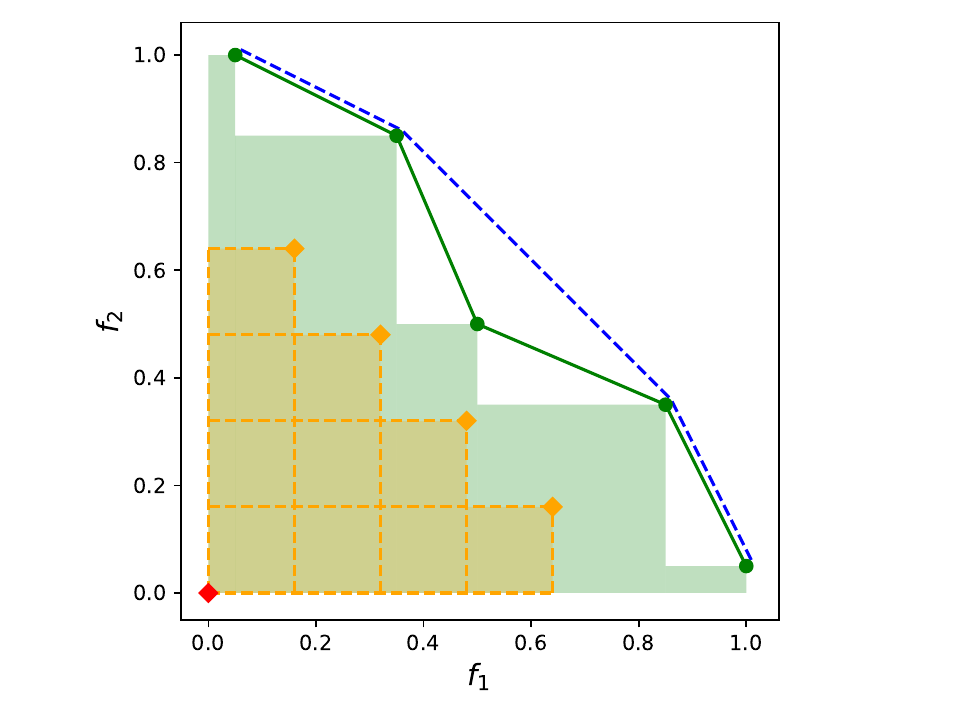}
    \caption{Pareto front \& HV for two objective functions: The red diamond represents the reference point $r$. The green points on the green solid line correspond to the optimal Pareto front, the green shaded area represents the optimal HV with respect to $r$, and the blue dashed line corresponds to the convex hull of the Pareto front.    
    The green point in the middle of the Pareto front does not lie on the convex hull and illustrates a non-supported Pareto optimal solution, the four other green points lie on the convex hull and illustrate supported Pareto optimal solutions. The orange diamonds represent examples of dominated solutions approximating the Pareto front, with the orange area illustrating their corresponding HV.}
    \label{fig:hyper_volume}
\end{figure}

Since the Pareto front and its approximations consist of a set of points, we cannot use the objective values directly to compare the quality of different given approximations. There exist many possible performance metrics for MOO with the hypervolume (HV) being the one most commonly used. A survey of performance metrics for MOO is given in \cite{riquelme_2015_moo_metrics}. The HV of a given set $\{(f_1(x), \ldots, f_m(x)) \mid x \in S\}$, where $S \subset X$ is a set of (unique) solutions, measures the volume spanned by the union of all hyperrectangles defined by a reference point $r \in \mathbb{R}^m$ and the points $(f_1(x), \ldots f_m(x))$, where $r$ is assumed to be a component-wise lower bound for all (given) vectors of objective values. 
If $S$ equals the Pareto front, the HV is maximized. The reference point $r$ can be chosen, e.g., by minimizing all objective functions individually. Then $r$ satisfies $r_i \leq f_i(x)$ for all $i \in \{1, \ldots, m\}$ and $x \in X$. 
Evaluating the HV exactly is in general \textbf{\#P}-hard in the number of dimensions. However, efficient approximation schemes exist, e.g., leveraging Monte Carlo sampling \cite{Fonseca_2006_hypervolume, Bringmann_2010, deng_2020_hv_mc}. 
Fig.~\ref{fig:hyper_volume} illustrates a non-convex Pareto front, its convex hull, (non-)dominated and \mbox{(non-)supported} solutions, as well as the optimal HV and the HV of a given set of solutions approximating the Pareto front.

\section{Classical Algorithms for MOO \label{sec:classical_algorithms}}

There are different classical strategies to approximate the Pareto front. The two most common approaches are the \emph{weighted sum method} (WSM) and the \emph{$\epsilon$-constraint method} ($\epsilon$-CM)~\cite{Ehrgott2005}. 
Both convert MOO into many single-objective optimization problems that are solved with conventional (combinatorial) optimization solvers.

The WSM constructs a single objective function
\begin{eqnarray}
    f_c(x) &=& \sum_{i=1}^m c_i f_i(x),
\end{eqnarray}
where $c \in [0, 1]^m$ with $\sum_i c_i = 1$ denotes a vector of weights of the convex combination of objectives. Solving $\max_{x \in X} f_c(x)$ for an appropriate discretization of $c$ vectors leads to solutions $x_c^*$ with corresponding objective vectors $(f_1(x_c^*), \ldots, f_m(x_c^*))$ and allows to approximate the Pareto front.
Assuming we can solve these problems to global optimality, e.g., by using commercial solvers like CPLEX \cite{ibm_cplex} or GUROBI \cite{gurobi}, all resulting points lie on the Pareto front.
However, the WSM can only find non-supported solutions, i.e., solutions whose corresponding vectors of objective values lie on the convex hull of the Pareto front \cite{Ehrgott2005}. As illustrated in Fig.~\ref{fig:hyper_volume} and shown in Appendix~\ref{sec:classical_results}, there can be a significant gap between the HV achieved by supported solutions and the optimal HV of all solutions, including non-supported ones.
In case of QUBO objective functions, the WSM results in a new QUBO for each $c$ and we denote the corresponding cost matrix by $Q_c = \sum_i c_i Q_i$.

\begin{figure}[ht]
    \centering
    \includegraphics[width=\linewidth]{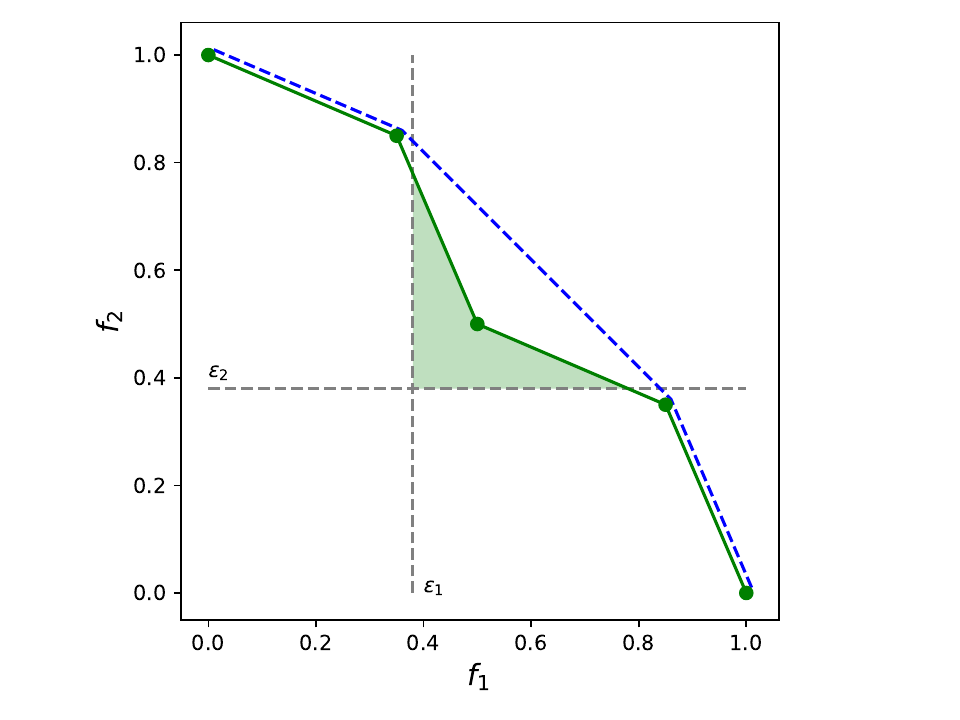}
    \caption{The $\epsilon$-CM: Given the lower bounds $\epsilon_i$ on the objective function values, the resulting feasible set (green area) only contains the non-supported Pareto optimal solution, which -- in this simplified example -- will be found by maximizing any of the two objective functions or convex combinations of them.}
    \label{fig:epsilon_constraints}
\end{figure}

The standard $\epsilon$-CM constructs single objective optimization problems by maximizing only one $f_i(x)$ at a time subject to constraints $f_j(x) \geq \epsilon_j$, $j \neq i$, on the other objective functions:
\begin{eqnarray}
    \max_{x \in X} && f_i(x)\\
    \text{subject to:} && f_j(x) \geq \epsilon_j, \, \forall j \neq i.
\end{eqnarray}
By solving such problems for all objective functions with appropriate discretizations of the $\epsilon_j$, this method allows to approximate the whole Pareto front including non-supported solutions.
Thus, if the discretization of the $\epsilon_j$ is sufficiently fine, the method can compute the exact Pareto front.
While the $\epsilon$-CM is more powerful than the WSM, adding additional constraints can make the problem more difficult to solve.
For instance, adding constraints maps QUBO to quadratic \emph{constrained} binary optimization.
Fig.~\ref{fig:epsilon_constraints} illustrates the $\epsilon$-CM.
In the presented form, the $\epsilon$-CM may also generate dominated solutions.
However, more advanced variants can overcome this inefficiency. 
For instance, one can combine the ideas of the WSM and the $\epsilon$-CM in a hybrid method~\cite{Ehrgott2005}.

For both methods, the number of $c$, respectively, $\epsilon$ discretization points to find good approximations of the Pareto front, in general, scales exponentially with the number of objective functions considered. Thus, they are only practical for a very small number of objective functions.
Therefore, we consider randomized variants of both that can also approximate the Pareto front for larger numbers of objective functions, as discussed in the following.

Instead of systematically discretizing the $c$ vector in the WSM, we propose to sample it randomly and solving the resulting problems with an appropriate solver. We sample them uniformly at random from the feasible set, cf.~Appendix \ref{sec:simplex_uniform_sampling} for more details.
This sampling-based approach allows us to approximate the efficient frontier without having to enumerate an exponential number of grid points.
As before, every sampled $c$ leads to a supported solution $x_c^*$ that lies on the Pareto front.
Alternative variants of the WSM, e.g., using the Goemans-Williamson algorithm \cite{goemans_williamson_1995_maxcut} to solve the resulting problems are possible, cf.~Appendix~\ref{sec:classical_results}.

Similarly, we propose a randomized variant of the $\epsilon$-CM. First, we minimize and maximize the individual objective functions to get lower and upper bounds $l_i \leq f_i(x) \leq u_i$, for all $x \in X$, where the lower bounds coincide with the reference point $r$ that we use for the HV, cf.~Sec.~\ref{sec:moo}. Then, we sample $(\epsilon_1, \ldots, \epsilon_m)$ uniformly at random from $\bigotimes_{i=1}^m [l_i, u_i]$. Instead of optimizing a particular objective $f_i$, we take a sample $c$ like in the randomized WSM and solve the problem
\begin{eqnarray}
    \max_{x \in X} && \sum_{i=1}^m c_{i} f_i(x)\\
    \text{subject to:} && f_j(x) \geq \epsilon_j, \, \forall j.
\end{eqnarray}
This method is a bit more complex than the randomized WSM due to the additional constraints, but it has two important advantages: 
First, the randomized $\epsilon$-CM is guaranteed to approximate the optimal Pareto front in the limit of infinitely many samples. Second, the ratio of feasible $\epsilon$ samples, i.e., samples where the resulting constrained optimization problem has a feasible solution, multiplied by the volume of the space of possible values $\prod_{i=1}^m (u_i - l_i)$ is a Monte Carlo estimator for the optimal HV. Both statements assume that the resulting mixed integer programs (MIPs) are solved to global optimality or that infeasibility can be proven if no feasible solution exists.

In recent years, numerous more involved algorithms have been proposed for specifically solving multi-objective integer problems for an arbitrary number of objectives. Many objective-space algorithms enumerate all solutions by decomposing the search space and applying a scalarization to the resulting subproblems \cite{Kirlik2014, Boland2017, Daechert2024} or by recursive reduction \cite{Ozlen2014}. In contrast, there are also decision-space algorithms, for example, following a branch-and-bound approach \cite{Bauss2023}.

Definitions of multi-objective integer programming vary across papers regarding the nature of the objective coefficients. While some algorithms are only proven to be correct for integer objective coefficients - such as DPA-a~\cite{Daechert2024} and AIRA~\cite{Ozlen2014} - others, in theory, can also handle continuous objective coefficients, including Epsilon~\cite{Kirlik2014}, DCM~\cite{Boland2017}, and Multi-Objective Branch-and-Bound~\cite{Bauss2023}. However, publicly available implementations of these algorithms are generally restricted to integer objective coefficients. Applying those algorithms to problems with continuous objective coefficients by scaling the objectives is possible, but not in all cases efficient due to a significant increase in the size of the objective space, which corresponds to the precision of the coefficients, leading to longer runtimes. As a representative for those multi-objective integer algorithms, we consider DPA-a and DCM.
See Appendix \ref{sec:classical_results} for further details.

In addition to the MIP-based algorithms mentioned above, there are also several genetic algorithms that have been proposed for MOO, see~eg.~\cite{deb_2002_nsga2}.
Getting these algorithms to work is known to require a lot of fine tuning of all steps involved. 
A first test of available implementations \cite{pymoo} did not lead to competitive results and we leave it to future work to further study these approaches.

\section{Quantum Approximate Optimization Algorithm\label{sec:qaoa}}

The QAOA~\cite{farhi2014quantumapproximateoptimizationalgorithm, farhi2015quantumapproximateoptimizationalgorithm} considers a QUBO problem on $n$ variables with cost matrix $Q$ and maps it to the problem of finding the ground state of a corresponding diagonal Ising Hamiltonian $H_C$ operating on $n$ qubits, which, for example, can take the form
\begin{equation}
    H_C = -\sum_{i<j} J_{ij}\sigma_i^z \sigma_j^z -\sum_i h_i \sigma_i^z,
\end{equation}
where $\sigma_i^z$ is the Pauli-Z operator acting on the $i$-th qubit, and $h_i$ and $J_{ij}$ are the problem coefficients. The encoding is achieved by representing the binary variables $x \in \{0, 1\}^n$ by spin variables $z \in \{-1, 1\}^n$, via the linear transformation $x_i=\frac{1-z_i}{2}$, and replacing the variables $z_i$ by $\sigma_i^z$ \cite{Lucas_2014}. This mapping leads to $\braket{x|H_C}{x} = - x^T Q x$ (up to a constant energy offset), where $\ket{x}$ is the quantum state encoding the bitstring $x\in \{0,1\}^n$. Thus, finding the ground state $\min_{\ket{x}} \braket{x|H_C}{x}$ corresponds to maximizing the objective function $x^T Q x$. This Hamiltonian $H_C$  that encodes the cost of the objective function is typically known as the \textit{phase separator} Hamiltonian. 

To solve the problem, QAOA first prepares the ground state of a \textit{mixer}  Hamiltonian, often conveniently chosen to be $H_X = -\sum_i \sigma_i^x$ with its ground state $\ket{+} = \frac{1}{\sqrt{2^n}}\sum_x \ket{x}$, where $\sigma_i^x$ is the Pauli-X operator acting on the $i$-th qubit. Then, QAOA applies alternating simulation of the two non-commuting Hamiltonians $H_X$, $H_C$, resulting in the state
\begin{eqnarray}
    \ket{\psi(\beta, \gamma)} &=& \prod_{k=1}^p e^{-i\beta_{k} H_X}e^{-i\gamma_{k} H_C} \ket{+},
\end{eqnarray}
with parameters $\beta, \gamma \in \mathbb{R}^p$, and number of repetitions $p \in \mathbb{N}$.
As $p$ approaches infinity, QAOA is guaranteed to converge to the ground state of the problem Hamiltonian~\cite{farhi2014quantumapproximateoptimizationalgorithm}, assuming the parameters $\beta$ and $\gamma$ are set appropriately. 
For finite $p$, carefully setting the parameters $\beta$ and $\gamma$ is crucial for the performance of the algorithm. 
The standard approach for selecting QAOA parameters uses a classical optimizer to minimize $\braket{\psi(\beta, \gamma)|H_C}{\psi(\beta, \gamma)}$ with respect to $(\beta, \gamma)$ in an iterative manner, where the objective function is evaluated using the quantum computer. 
Once good parameters have been found, sampling from $\ket{\psi(\beta, \gamma)}$ results in good solutions to the original QUBO.
There exist several variants of QAOA, such as recursive QAOA \cite{Bravyi_2020} or warm-started QAOA \cite{Egger2021, tate2022bridgingclassicalquantumsdp}, which can help to further improve the performance. However, within this paper we focus on the original proposal.

The highly non-convex parameter landscape, sampling errors, and hardware noise make it challenging to find good parameters \cite{rajakumar2024trainability}.
However, there exist multiple alternative strategies to find good parameters for QAOA and circumvent the trainability challenge.
For example, as proposed in \cite{Streif2020} the expectation values for training QAOA parameters can be evaluated classically. Then, the quantum computer is only used to generate the samples. This can be of potential interest since, for QUBO, $H_C$ consists of only 1- and 2-local terms, which can be significantly cheaper to evaluate classically than sampling from the full quantum state.
Alternatively, good angles can be derived for certain problem classes from an infinite problem size limit \cite{Farhi2022, Boulebnane2021, Basso2022, PhysRevA.103.042612, wurtz2021fixedangleconjectureqaoa}, by applying linear parameter schedules \cite{Montanez-Barrera2024}, or by using machine learning models \cite{Khairy2020, Alam2020}.
Another promising strategy is to reuse optimized parameters from similar problem instances \cite{montanezbarrera2024transferlearningoptimalqaoa, 9605328, augustino2024strategiesrunningqaoahundreds}. This is often referred to as \textit{parameter transfer} or \emph{parameter concentration} and is discussed in more detail in the following.

Parameter transfer in QAOA is a method to reuse the parameters obtained for one problem instance for another one without the need for repeating the classical parameter optimization.
Optimized QAOA parameters for one problem instance often result in high-quality solutions for another instance, if both problem instances exhibit similar properties. 
This has been proven analytically on low-depth MAXCUT problems on large 3-regular graphs \cite{Brandão2019} and observed empirically on QAOA with $p=1$ for MAXCUT problems on random regular graphs with same parity \cite{Galda2023}. Strategies to rescale the unweighted MAXCUT parameters for weighted problems exist to enable parameter transfer \cite{Sureshbabu2024, Shaydulin2023}. Several studies show parameter concentration of similar problem instances with respect to problem size on random Hamiltonians empirically and analytically \cite{Streif2020,Akshay2021,pelofske2023short,Pelofske_2024,augustino2024strategiesrunningqaoahundreds}. In general QAOA parameter transfer is a good heuristic tool to efficiently find QAOA angles which perform reasonably well, and importantly can circumvent computationally costly variational learning. Importantly, parameter transfer does not yield optimal QAOA angles in general for all instances, but instead can work well on average for ensembles of similarly structured Hamiltonians. 
Parameter transfer is also possible from smaller to larger problem instances, and different training strategies can be combined, e.g., we can use some of the established schemes to initialize the parameters for a smaller problem, then train it using classical evaluation of the expectation values, and transfer the resulting parameters to larger problem instances with similar structure. This is the direction we will be applying in the following.
A broad study on parameter transfer for different problem classes has been conducted in \cite{Katial2024InstanceQAOA}.
In the following, we show how to leverage QAOA and parameter transfer in the context of weighted MAXCUT MOO on specific quantum computer hardware graphs.

\section{QAOA for Multi-Objective Optimization}
\label{sec:quantum_moo}

There exists only very little literature on quantum algorithms for MOO. 
Ref.~\cite{Chiew_2024} proposes to use QAOA as a solver in the WSM, similar to what we will propose here. Within the WSM, a weight vector $c$ maps multiple QUBO cost matrices $Q_i$ to a new combined QUBO cost matrix $Q_c$. Thus, one can run QAOA for every $c$ vector independently. In \cite{Chiew_2024}, the authors propose to re-optimize the QAOA parameters $(\beta, \gamma)$ for every $c$, which is computationally expensive and leads to a bottleneck.
Simlarly, Ref.~\cite{zakaria_2024_mo_qaoa} suggests to re-optimize the QAOA parameters. However, the authors suggest to warm-start every optimization with the parameters found in the previous iteration and motivate it by QAOA parameter transfer. In addition, they propose a Tchebycheff scalarization strategy to navigate the Pareto front.
In contrast, \cite{ekstrom2024variational} propose to directly use the HV as a single objective function in a variational quantum algorithm. Further, they propose an extended ansatz that includes QAOA cost operators for every objective function in the considered MOO problem.
This can be potentially problematic, as one tries to generate a state that corresponds to a superposition of Pareto optimal solutions and thus can become more difficult to train in cases where the Pareto front consists of many points.
Further, they extend the QAOA ansatz by including a cost operator for every objective function, which is not covered by any of the existing theory on the performance of QAOA and leads to deeper circuits.
Lastly, \cite{li2024distributedexactmultiobjectivequantum} introduce an algorithm based on Grover's search \cite{grover1996fastquantummechanicalalgorithm} with a quadratic speed-up over brute force search leading to large quantum circuits that require error correction.

We propose to go one step further than Refs.~\cite{Chiew_2024, zakaria_2024_mo_qaoa}and completely skip the optimization of the QAOA parameters for the considered problem instance. 
More precisely, in order to overcome the aforementioned computational bottleneck introduced by re-optimizing for every $c$ vector, we leverage the transfer of parameters discussed in Sec.~\ref{sec:qaoa} from a single (smaller) problem instance to the (larger) ones resulting for every sampled $c$ vector. Thus, we only have to train the QAOA parameters once, which can be done offline, i.e., before the actual problem instance of interest is known.
For this to work, we have to identify a representative single-objective problem and find good QAOA parameters for it.
For the training, we use problems small enough to be simulated classically, i.e., there is no quantum computer needed for the parameter training. 
We then fix these QAOA parameters and only vary the cost Hamiltonian through sampling $c$ vectors.
The rationale is that QAOA with fixed parameters gives a diverse set of good solutions for all $c$ vectors, although not necessarily the optimal solutions. However, a sub-optimal solution for a particular $c$ vector might still be a Pareto optimal solution, and since it does not need to be optimal for the considered weighted sum, it might also result in non-supported Pareto optimal solutions. 

As mentioned in Sec.~\ref{sec:qaoa}, transfer of parameters can apply between similar instances of the same size, but also from smaller to larger instances. Here, we propose to train parameters classically on sufficiently small but representative instances and then re-use them for larger instances on a quantum computer. Since the parameters are considered independent of the concrete problem instance and can be determined offline, the training does not contribute to the runtime of an algorithm for a particular problem instance and we only consider the actual sampling from the QAOA circuits.

This approach of using QAOA to sample sub-optimal solutions to the underlying diagonal Hamiltonian cost function with the goal of sampling the Pareto optimal front is notable for several reasons: First, this means that we need both high-diversity solution sampling of configurations that share similar cost values--i.e.~fair sampling \cite{fair_sampling_NISQ, fair_sampling_NISQ_pelofske}--and high-variance sampling which specifically means we do not use QAOA circuits converged to the ground-state (i.e., approximation ratio of $1$, which again would rule out non-supported solutions). Second, which follows from the first reason, we utilize low depth QAOA circuits, which makes the computation especially suitable for noisy quantum processors. 
Last, there are several no-go theorems on QAOA, usually providing lower bounds on the number of QAOA repetitions $p$ with respect to the problem size to achieve on average a certain solution quality \cite{Bravyi_2020, farhi2020quantumapproximateoptimizationalgorithm}. However, for the considered algorithm, none of the known no-go theorems apply, since a solution that might be bad for a particular $c$ vector may still be (non-supported) Pareto optimal. 

In the following, we demonstrate the potential of our proposal on a concrete problem and show that QAOA with transfer of parameters represents a promising approach for MOO.

\section{Experimental Results}
\label{sec:experiments}

We demonstrate our proposal on instances of MO-MAXCUT. The goal of single-objective weighted MAXCUT is to partition the set of nodes $\mathcal{V}$ of a given weighted graph $\mathcal{G} = (\mathcal{V}, \mathcal{E}, w)$, with edge weights $w_{kl}$ for edges $(k, l) \in \mathcal{E}$, into two sets, maximizing the sum of the edge weights of the edges connecting the two sets. Weighted MAXCUT can be formulated as a QUBO:
\begin{eqnarray}
    \max_{x \in \{0, 1\}^n} \sum_{(k, l) \in \mathcal{E}} w_{kl} (x_k + x_l - 2 x_k x_l).
\end{eqnarray}
To introduce multiple objective functions, we assume multiple weighted graphs $\mathcal{G}_i = (\mathcal{V}, \mathcal{E}, w^i)$, $i = 1, \ldots, m$, defined on the same set of nodes $\mathcal{V}$, and here, for simplicity, also on the same set of edges $\mathcal{E}$, i.e., they only differ in their edge weights $w^i$. Then, each graph defines a MAXCUT objective function over the same decision variables.

\begin{figure}
    \centering
    \includegraphics[width=1\linewidth]{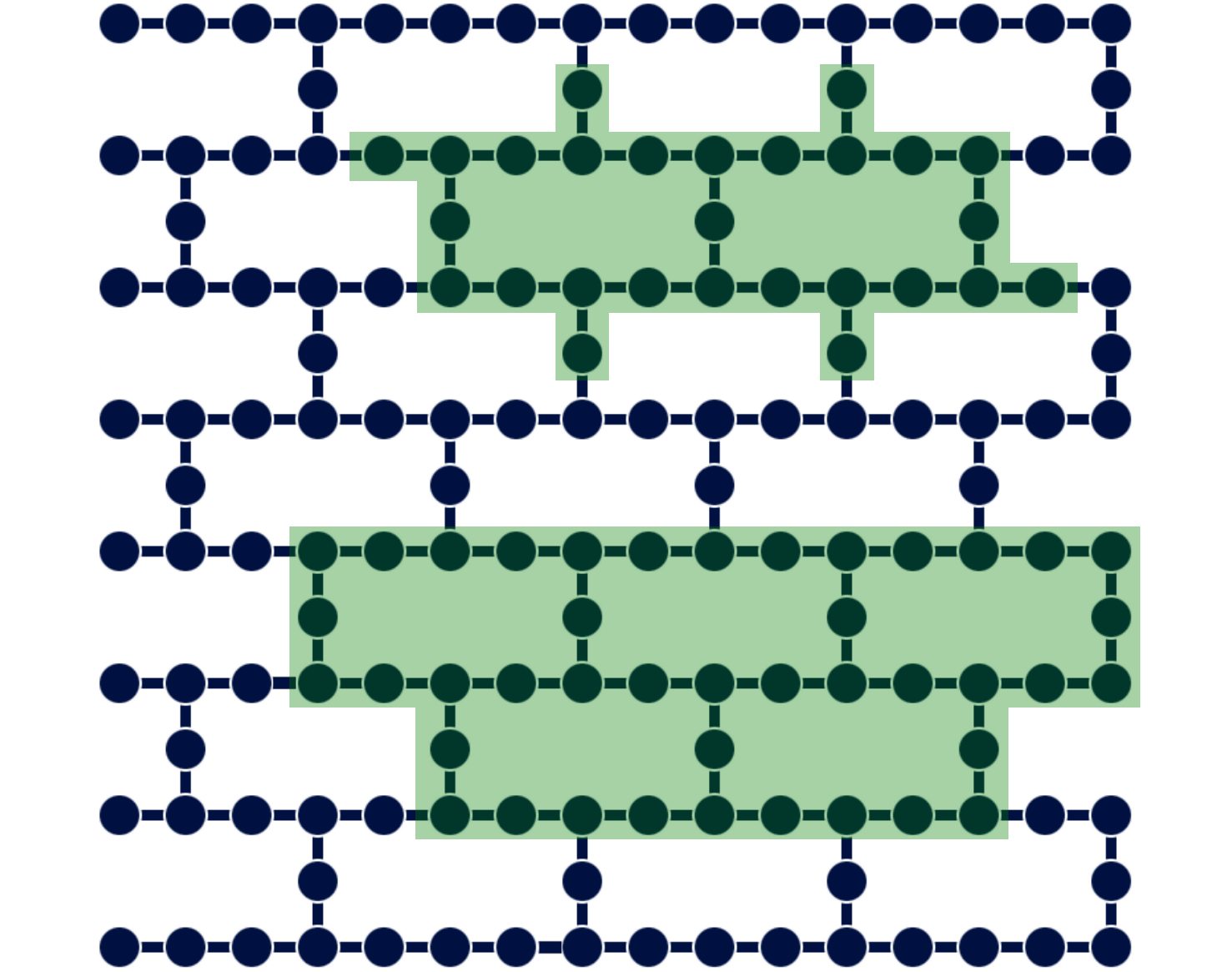}
    \caption{Graph topologies: The whole graph shows the coupling map of the 156-qubit \emph{ibm\_fez} device. The green-shaded areas show the 27-qubit graph of the training problem (top) and the 42-qubit graph of the considered MO-MAXCUT problem (bottom) embedded in the \emph{ibm\_fez} coupling map.}
    \label{fig:topology}
\end{figure}

We consider subgraphs of the heavy-hex topology \cite{Chamberland_2020} of IBM Quantum devices \cite{ibm_quantum_platform} to facilitate execution on real quantum hardware, see Fig.~\ref{fig:topology}.
Defining the problem instance using the native two-qubit gate connectivity graph of the hardware allows to construct relatively short-depth QAOA circuits that can be executed with high fidelity~\cite{romero2024biasfielddigitizedcounterdiabaticquantum, kim2023evidence, pelofske2023short, Pelofske_2024, QAOA_vs_QA}.
More precisely, we consider 42-node graphs to define the MO-MAXCUT problem with $m=3$ and $m=4$ objective functions.
To implement the corresponding circuits on hardware, we need a parametrized RZZ-gate for each edge in the graph for the implementation of the phase separator cf.~Sec.~\ref{sec:qaoa}.
These gates are then compiled to the hardware native basis gate set \{$X$, $\sqrt{X}$, RZ, CZ\}, i.e., a RZZ-gate consists of two CZ-gates plus single qubit gates.
The minimum edge-coloring of the sparse heavy-hex lattice requires three colors, thus, we can apply all RZZ-gates in three non-overlapping layers. This determines the RZZ-depth as three for each QAOA round and the corresponding CZ-depth as six, assuming RZZ-gates are implemented using two CZ-gates. 
Since, on the considered device sub-graph that we compile the QAOA circuits to, all RZZ-gates have a uniform duration of 84ns, the choice of edge coloring does not affect the total compiled circuit duration and therefore a random 3-edge-coloring is used. The mixing Hamiltonian is implemented using single qubit RX-gates per layer compiled to the basis gates. 

For each graph, the edge weights are independently sampled from a standard normal distribution, i.e., $ w_{kl}^i \sim \mathcal{N}(0, 1)$, for $(k, l) \in \mathcal{E}$, $i = 1, \ldots, m$. 
Note that considering continuous weights make these problems particularly challenging for classical algorithms, cf.~Sec.~\ref{sec:classical_algorithms} and Appendix~\ref{sec:classical_results}.

Further, we consider a 27-node graph for the classical QAOA parameter optimization with weights sampled independently from $\mathcal{N}(0, 1/m)$ for the different number of objective functions.
The standard deviation $1/m$ accounts for the randomized convex combinations of the $m$ objective functions used in our algorithm.
For MO-MAXCUT, this corresponds to convex combinations of the edge weights $w^i$, leading to a lower standard deviation compared to edge weights drawn form the standard normal distribution. The standard deviation of $1/m$ is equivalent to taking the average of $m$ random weights drawn independently from a standard normal distribution.
The corresponding (unweighted) 27-node and 42-node graphs are shown in Fig.~\ref{fig:topology}.
The structure of the 27-node graph was inspired by the topology of the 27-qubit IBM Quantum Falcon devices. However, in the present context we simulate the corresponding circuit classically for initial parameter training and only execute the 42-qubit circuits on a quantum computer for generating the results of the problem of interest.

To optimize the QAOA parameters for the 27-node single-objective MAXCUT problem, we use~\texttt{JuliQAOA} \cite{Golden_2023}, which is a Julia-based \cite{bezanson2017julia} QAOA simulator. It efficiently evaluates expectation values and corresponding gradients based on the QAOA-statevector and applies a basin-hopping optimization algorithm to navigate the many local minima in the parameter landscape. More details on the optimization pipeline are provided in Appendix~\ref{sec:juli_qaoa}.
We empirically confirm that transfer of parameters works for the considered class of problems, as illustrated in Appendix~\ref{sec:transfer_of_parameters}.
This allows us to pre-optimize the QAOA angles for $p=1, \ldots, 6$, independently of the concrete 42-node MO-MAXCUT problem instance.
In the following we discuss the results for the two considered problem instances.

\subsection{MO-MAXCUT with 3 objective functions}
\label{sec:experiments_3o}

To asses and compare the performance, we evaluate the progress of the achieved hypervolume $\text{HV}_t$ over time $t$. We always report single-thread CPU and QPU times.
First, we tested the classical algorithms discussed in Sec.~\ref{sec:classical_algorithms} on the 42-node MO-MAXCUT problem with $m=3$ objective functions.
DCM, DPA-a, and the $\epsilon$-CM performed best and are discussed in more detail here. For the results of the other algorithms see Appendix~\ref{sec:classical_results}.

For DCM and DPA-a, we scaled the continuous weights by a factor of \num{1000} and rounded them to integers, since the available implementations of the algorithms cannot handle continuous weights natively. 
While we set a time limit of \num{10000}s for both algorithms, 
DPA-a terminated already after 3.6min and DCM after 8min. Thus, both find the global optimal hypervolume for the truncated weights.
The HV achieved by the resulting non-dominated points evaluated for the exact non-truncated weights is 
$\text{HV}_\text{max} = \num{43471.704}$.
Thus, this is the optimal HV up to the error introduced by the truncation of weights.
Both algorithms found 2063 non-dominated solutions.

Note that particularly DCM was very sensitive to the scaling factor. When the weights were scaled by a factor of \num{10000} before rounding, the algorithm only found as little as four non-dominated points before reaching the given time limit. This shows that continuous weights can pose a particular challenge for classical algorithms.
DPA-a was more robust to the scaling factor for the considered problem instance and achieved a comparable performance in both cases, cf.~Sec.~\ref{sec:classical_results}.

\begin{figure*}[ht!]
    \centering
    \includegraphics[width=1\linewidth]{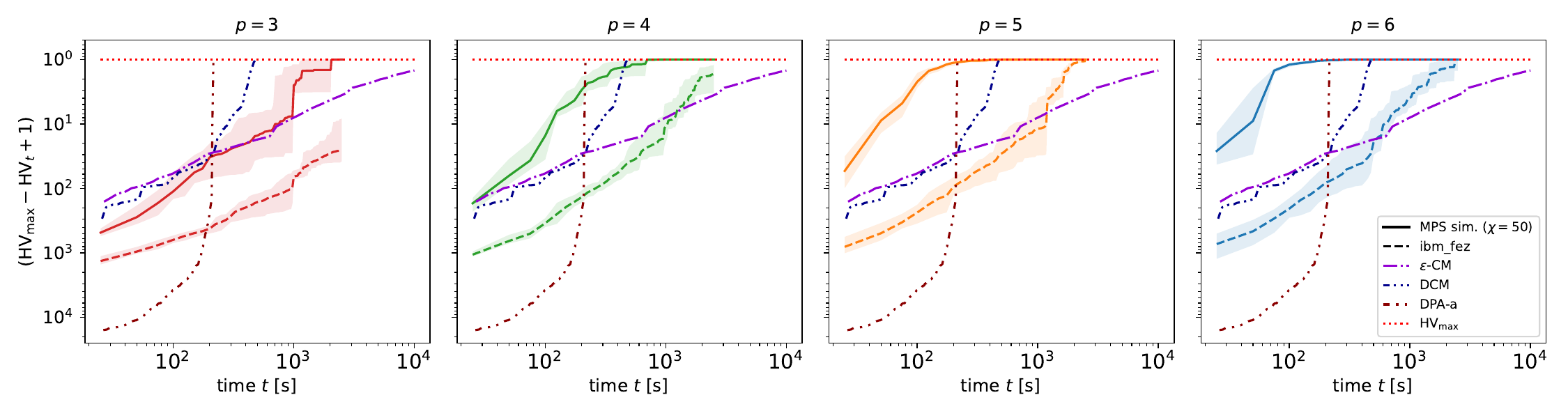}
    \caption{Comparison of quantum and classical results for MO-MAXCUT with $m=3$ objective functions: Average progress of $\text{HV}_t$ of the MPS simulation ($\chi = 50$, solid line), \emph{ibm\_fez} (dashed line), and the classical algorithms DCM, DPA-a, and $\epsilon$-CM (dotted-dashed lines). The shaded areas denote the minimum and maximum performance over the five repetitions of the MPS simulation and results from \emph{ibm\_fez}. The best found solution $\text{HV}_{\text{max}}$ is indicated by the red dotted horizontal line.}
    \label{fig:42q_hw_results}
\end{figure*}

\begin{figure}
    \centering
    \includegraphics[width=1\linewidth]{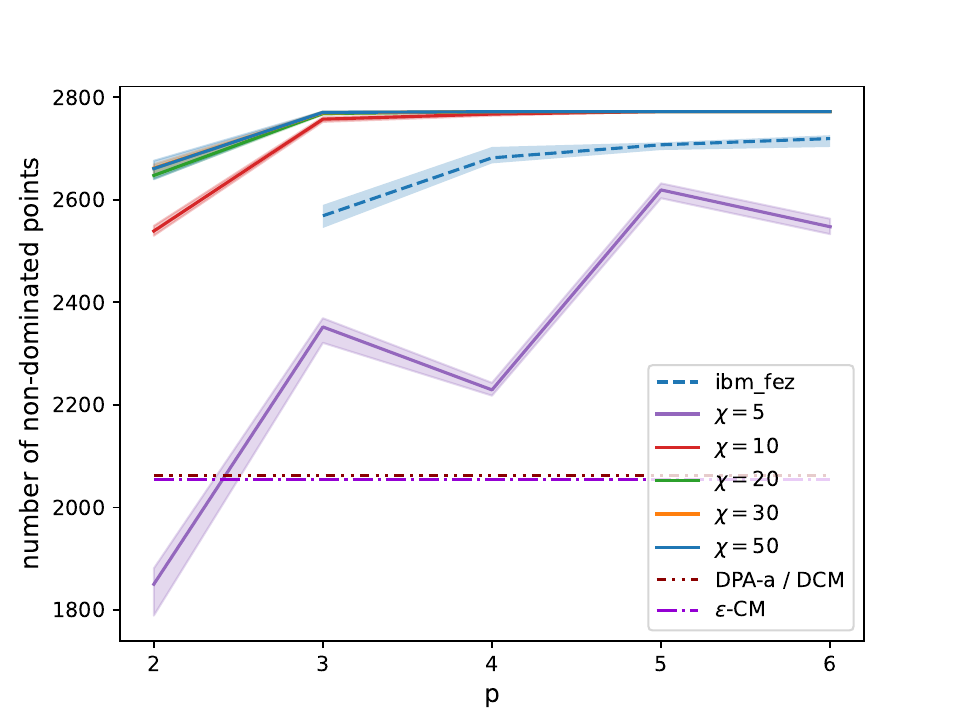}
    \caption{Number of non-dominated points: The solid lines show the average number of unique non-dominated points approximating the Pareto front resulting from the five repetitions using MPS simulation, with the shaded areas denoting the corresponding minimum and maximum. The largest value reached is \num{2772}. For $\chi = 1$ or $p=1$, the number of points was only between 170 and 255 and is not shown here. The dashed line and corresponding shaded area shows the results from the runs on \emph{ibm\_fez} reaching a maximum of \num{2726}. For both, the MPS simulation and the results from \emph{ibm\_fez} 25 million shots were taken in total.
    The dotted-dashed lines show the result for DPA-a and DCM (both achieve the same value) and the $\epsilon$-CM.}
    \label{fig:42q_num_non_dominated_points}
\end{figure}

The HV-progress over time for all algorithms discussed in this section is shown in Fig.~\ref{fig:42q_hw_results}.
In all figures showing the HV progress, we plot $(\text{HV}_\text{max} - \text{HV}_t + 1)$ in a log-log plot with a reversed y-axis to increase the contrast in the visualization while preserving the qualitative presentation of the results.
The number of non-dominated points for all tested algorithms is shown in Fig.~\ref{fig:42q_num_non_dominated_points}.

Next, we executed $\epsilon$-CM, again setting a time limit of \num{10000}s.
In our tests, every resulting MIP has been solved to optimality using GUROBI on average in \num{0.0218}s using a single-thread on \emph{AMD EPYC 7773X} CPUs. Thus, we sampled almost \num{460000} random $\epsilon$ and $c$ vectors and solved the corresponding MIPs.
We based our implementation on the standard integer programming formulation of MAXCUT.
Since the randomized $\epsilon$-CM adds constraints to the different objectives this is more suitable for MIP solvers than the QUBO formulation, which would result in non-convex quadratic constraints. See Appendix~\ref{sec:mip_maxcut} for more details.
The maximum HV achieved by the $\epsilon$-CM was \num{43471.222}, i.e., slightly below $\text{HV}_\text{max}$, where the Pareto front was approximated by \num{2054} non-dominated points. 

As discussed in Sec.~\ref{sec:classical_algorithms}, the ratio of feasible $\epsilon$ samples allows us to derive a Monte Carlo estimate of the optimal HV.
From the close to 460,000 samples, $84.917\%$ were feasible. The lower and upper bounds for the 3 objective functions are given by $l = (-12.137, -19.642, -18.331)$ and $u = (21.390, 19.130, 21.068)$, respectively. The volume of the resulting hyper rectangle is $V_{l,u} = \num{51212.547}$. Thus, the Monte Carlo estimate of the globally optimal HV is given by $\num{43488.364}$ with 95\%-confidence interval equal to $[\num{43435.378},\; \num{43541.351}]$, which contains and closely agrees with $\text{HV}_{\text{max}}$.

As a first test of the QAOA-based algorithm, we use Qiskit Aer \cite{qiskit_aer, javadiabhari2024quantumcomputingqiskit} to run Matrix Product State (MPS) simulations of the QAOA circuits with bond dimensions $\chi = 1, 5, 10, 20, 30, 50$ and $p = 1, \ldots, 6$. For each setting we draw \num{5000} random $c$ vectors as described in Appendix~\ref{sec:simplex_uniform_sampling}. Given the precomputed QAOA parameters from the 27-qubit problem instance, every sampled $c$ vector fully defines a corresponding QAOA circuit and we take \num{5000} shots from each of them. In total, this results in 25 million shots.

\begin{figure*}[h!t]
    \centering
    \includegraphics[width=\textwidth]{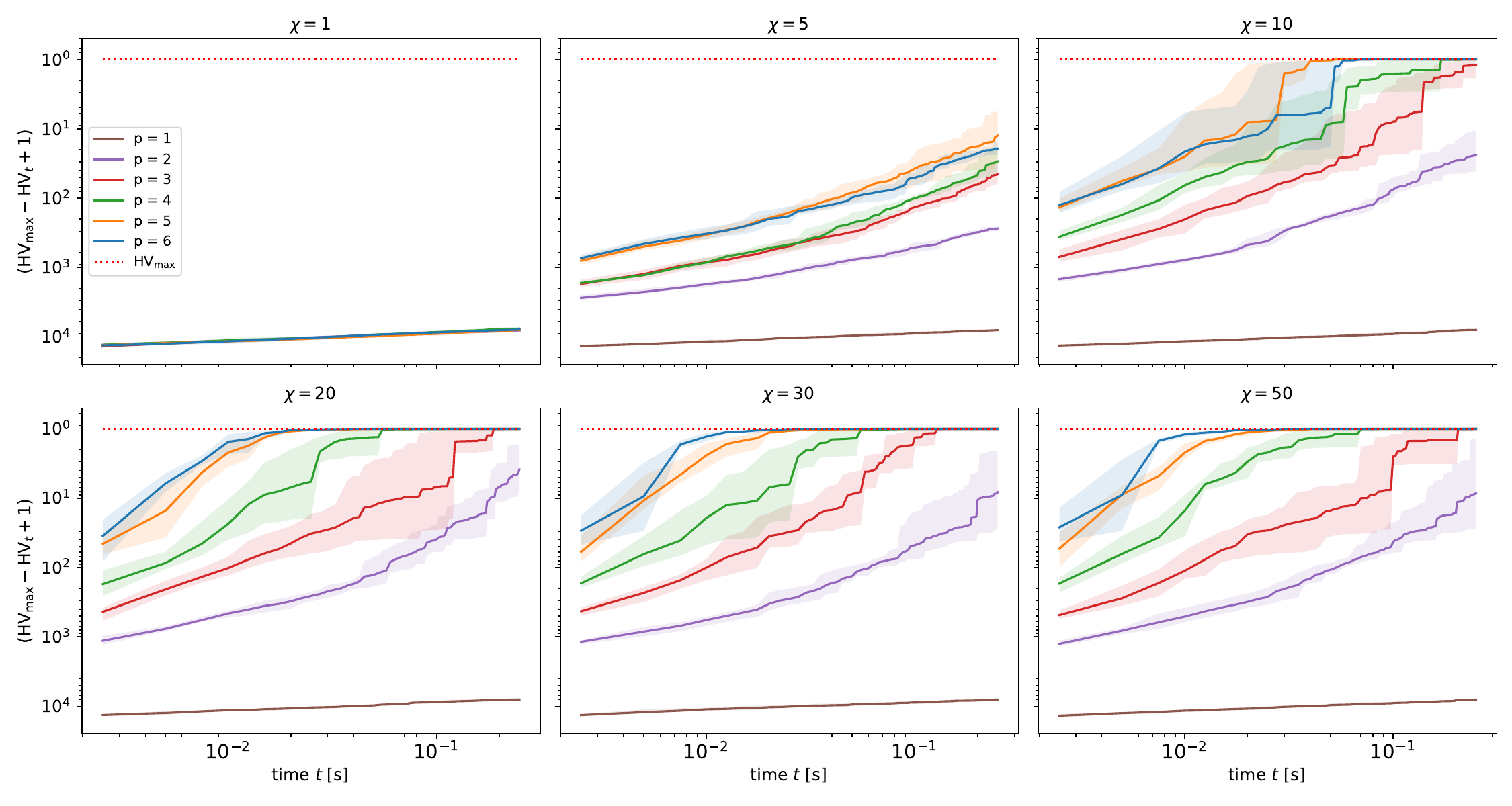}
    \caption{MPS simulation results: Hypervolume $\text{HV}_t$ for different bond dimensions $\chi$ and QAOA rounds (i.e., depth) $p$. Each line represents the average of five random repetitions of the same setting with shaded areas indicating the minimum and maximum values. The best found solution $\text{HV}_{\text{max}}$ is indicated by the red dotted horizontal line.}
    \label{fig:42q_mps_sim_results}
\end{figure*}

We are interested in the performance when running the circuits on a real quantum computer. Thus, we assume a sampling rate of \num{10000} shots per second, which is realistic as discussed later in the context of our hardware experiments, although the actual MPS simulations may take longer, cf.~Appendix~\ref{sec:mps_convergence}. 
Thus, 25 million shots are assumed to take in total \num{2500}s.
Then, we compute the corresponding hypervolume $\text{HV}_t$ at time $t$. 
With increasing $\chi$ and $p$, the simulations reach $\text{HV}_\text{max}$.
Each setting is simulated five times to analyze the robustness of the algorithm and we determine the resulting minimum, maximum, and average performance. The results are shown in Fig.~\ref{fig:42q_mps_sim_results}. 
A bond dimension $\chi = 20$ seems sufficient to evaluate the considered circuits, as there are only small difference between the results for $\chi = 20$, $30$ and $50$. This is further supported by the analysis presented in Appendix \ref{sec:mps_convergence}.
The resulting number of non-dominated points approximating the Pareto front is shown in Fig.~\ref{fig:42q_num_non_dominated_points}. It achieves a maximum of \num{2772}. The difference to the number of points found by DCM and DPA-a can be explained by the truncation of weights for the classical algorithms. However, this also implies that the additional points have only a negligible contribution to the HV.

Finally, we run the quantum algorithm for $p = 3, 4, 5, 6$ on the \emph{ibm\_fez} device accessed via the IBM Quantum platform \cite{ibm_quantum_platform}, again using \num{5000} random $c$ vectors and \num{5000} shots per circuit. 
There are twelve native options to place the considered circuits on the topology of \emph{ibm\_fez} without inserting any SWAP gates, one of them shown in Fig.~\ref{fig:topology}.
Before running the experiments, we estimate the resulting fidelity of running the circuit on the quantum computer using the error rates for gates and measurements reported by the backend.
Then we select the initial layout, i.e., subset of qubits, that leads to be highest fidelity.
The estimated fidelities are reported in Tab.~\ref{tab:circuit_summary} together with number of gates, depth, and duration of executing the circuits.
As in the MPS simulation, each experiment is repeated five times. 
The repetition delay between measuring one quantum circuit and starting the subsequent one is set to $10^{-4}$s. 
Since, in the considered case, executing a circuit is significantly shorter, see the durations reported in Tab.~\ref{tab:circuit_summary}, the repetition delay dominates the overall runtime on the quantum computer. This results in the aforementioned sampling rate of \num{10000} shots per second. Further reducing the delay may speed up the calculation but could reduce the quality by affecting the qubit initialization between shots.
Except for the adjusted repetition delay, we use the default execution pipeline provided by the IBM Quantum \emph{SamplerV2} primitive accessible via the IBM Qiskit Runtime \cite{ibm_quantum_platform}.
This sampling rate is an idealistic assumption, since it excludes, e.g., circuit compilation. However, we believe this is reasonable since all required compilation could be done ahead of time for a parametrized template that is independent of the concrete problem instance.

\begin{table}
\centering
\begin{tabular}{ccccc}
$p$ & \#CZ & depth (CZ/total) & duration [s] & fidelity [\%] \\
\hline
3   & 276  & 18 / 49          & $3.9\times 10^{-6}$ & 13.63 \\
4   & 368  & 24 / 65          & $4.7\times 10^{-6}$ & 8.84 \\
5   & 460  & 30 / 81          & $5.5\times 10^{-6}$ & 5.42 \\
6   & 552  & 36 / 97          & $6.3\times 10^{-6}$ & 3.71 \\
\end{tabular}
\caption{Summary of executed QAOA circuits: QAOA repetitions $p$, number of CZ gates, corresponding CZ depth, total gate depth, schedule duration, and estimated fidelity based on device calibration data. The duration of a CZ gate on \emph{ibm\_fez} uniformly is 84ns, which makes up for the majority of the circuit duration. The remainder is coming from single qubit gates and final measurements.}
\label{tab:circuit_summary}
\end{table}

Fig.~\ref{fig:42q_hw_results} shows the results for DPA-a, DCM, $\epsilon$-CM, the MPS simulations with $\chi = 50$, and the results from \emph{ibm\_fez}.
Assuming a 10kHz sampling rate, the MPS simulations with $p \geq 5$ outperformed all other methods and achieved $\text{HV}_{\text{max}}$ at about the same time as DPA-a.
All runs with $p \geq 3$ reached $\text{HV}_{\text{max}}$ within the \num{2500}s for all five repetitions.
The corresponding number of non-dominated points approximating the Pareto front are shown in Fig.~\ref{fig:42q_num_non_dominated_points}.
Notably, the samples generated by \emph{ibm\_fez} also improve with increasing $p$, i.e., with increasing circuit complexity, and for $p = 4, 5, 6$ overtake the $\epsilon$-CM on average after \num{1230}s, \num{1190}s, and \num{535}s, respectively. For $p=4$, the average results from the \emph{ibm\_fez} after \num{2500}s are comparable to the results of the $\epsilon$-CM after \num{10000}s. For $p= 5, 6$, the average results even overtake the best results from the $\epsilon$-CM and approach the best observed value $\text{HV}_{\text{max}}$.
This is quite remarkable and underscores the potential of the proposed approach.

The presented results allow already to draw two key conclusions:
First, they clearly show that performance improves with increasing bond dimension. While for $\chi = 1$, the circuit complexity $p$ has no impact, the performance greatly improves as $\chi$ increases, i.e., as the MPS simulator approximates the circuits more accurately.
Second, at larger bond dimensions, the performance consistently improves with increasing $p$, indicating that achieving good results requires executing sufficiently complex circuits with sufficiently high accuracy.
This is a strong indication that with increasing scale, this algorithm requires execution on a real quantum computer, in contrast to an approximation using, e.g., an MPS simulator.
These observations are consistent with the work of Ref.~\cite{Santra2025} that shows how genuine multi-partite entanglement is needed to obtain good results with quantum approximate optimization.
This provides a first strong indication that MOO is a promising candidate for a potential quantum advantage in optimization.
Last, for the problems considered here, with relatively small scale

\begin{figure}
    \centering
    \includegraphics[width=1\linewidth]{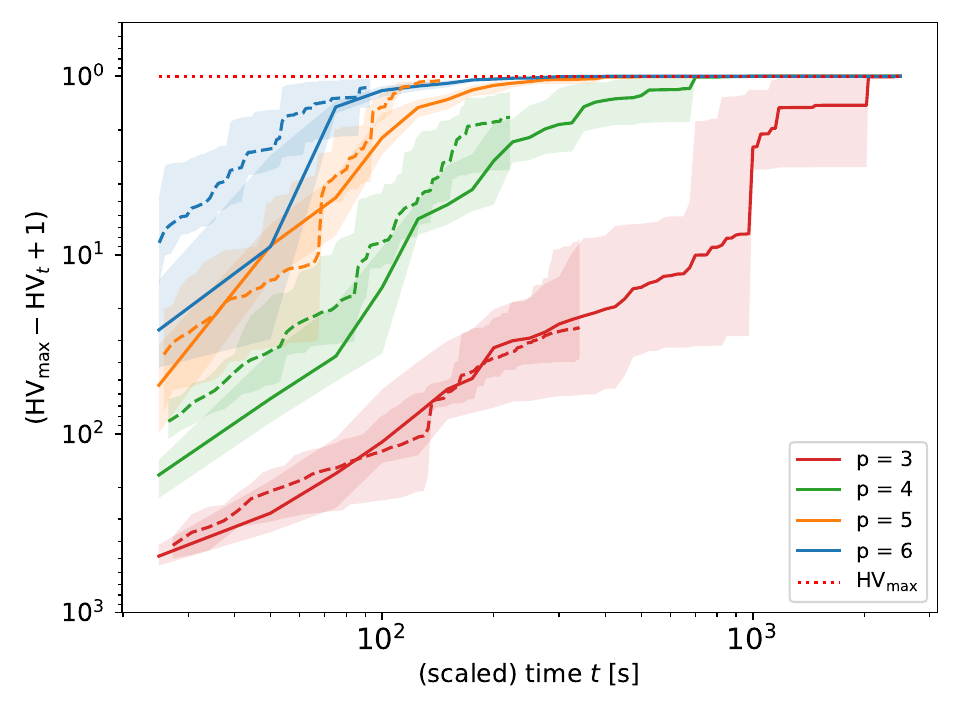}
    \caption{Fidelity-scaled quantum results: MPS simulation and \emph{ibm\_fez} results from Fig.~\ref{fig:42q_hw_results} where the time of the hardware results is scaled by the corresponding estimated fidelity reported in Tab.~\ref{tab:circuit_summary}.}
    \label{fig:42q_hw_results_scaled}
\end{figure}

The samples generated by the quantum computer are affected by noise, which explains the differences to the MPS simulations.
The estimated fidelities in Tab.~\ref{tab:circuit_summary}, derived from the errors reported by the backend, provide an indication of the strength of the noise. 
As discussed in~\cite{barron2024cvar}, the fidelity can be interpreted as the \emph{probability of no error}. 
This implies that on average a fraction equal to the fidelity of the generated samples are drawn from the ideal noise-free distribution, while the remaining samples are potentially corrupted by noise.
If, for simplicity, we assume that only samples corresponding to the noise-free distribution contribute to the HV and we discard the other ones, and if we scale the time it takes to generate the noisy samples from the quantum computer by the corresponding fidelity, the average $\text{HV}_t$ corresponding to the noise-free MPS simulation should result in a lower bound of the $\text{HV}_t$ corresponding to scaled hardware results. 
These scaled results, shown in Fig.~\ref{fig:42q_hw_results_scaled}, match the MPS simulation results very well.
This demonstrates how evaluating quantum heuristics on noisy devices can forecast their performance when being run on devices with lower error rates, including fault-tolerant devices, with a sampling overhead corresponding to the reciprocal of the fidelity.
In the following section, we will use this insight in the opposite direction for the analysis of the instance with $m=4$ objective functions.

\subsection{MO-MAXCUT with $m=4$ objective functions}
\label{sec:experiments_4o}

Next, we analyze a MO-MAXCUT problem instance with $m=4$ objective functions.
The underlying unweighted graph is the same as in Sec.~\ref{sec:experiments_3o} and the weights for the four objective functions are again sampled from $\mathcal{N}(0, 1)$.
As explained before, the weights for the 27-qubit training instance are sampled from $\mathcal{N}(0, 1/m)$ and the corresponding QAOA parameters are trained with JuliQAOA, cf.~Sec.~\ref{sec:transfer_of_parameters} for more details on the training and transfer of parameters.

As in Sec.~\ref{sec:experiments_3o}, we test DCM and DPA-a with weights scaled by a factor \num{1000} and then rounded to integers. Further, we run the $\epsilon$-CM. For all three algorithms we set a time limit of \num{20000}s, which results in about \num{740000} random $c$ and $\epsilon$ vectors--i.e.~MIPs--for the $\epsilon$-CM.
Since we have to add an additional constraint to the corresponding MIPs, the average time to solve one instance slightly increase to \num{0.0270}s per MIP compared to \num{0.0218}s for the case of three objective functions.
Again, DPA-a performs best among the tested classical algorithms but takes significantly longer than for three objectives. It terminates just before reaching the time limit and finds the optimal Pareto front for the truncated weights, with $\text{HV}_{\text{max}} = \num{1266143.350}$ when evaluated using the exact weights.

\begin{figure}
    \centering
    \includegraphics[width=1\linewidth]{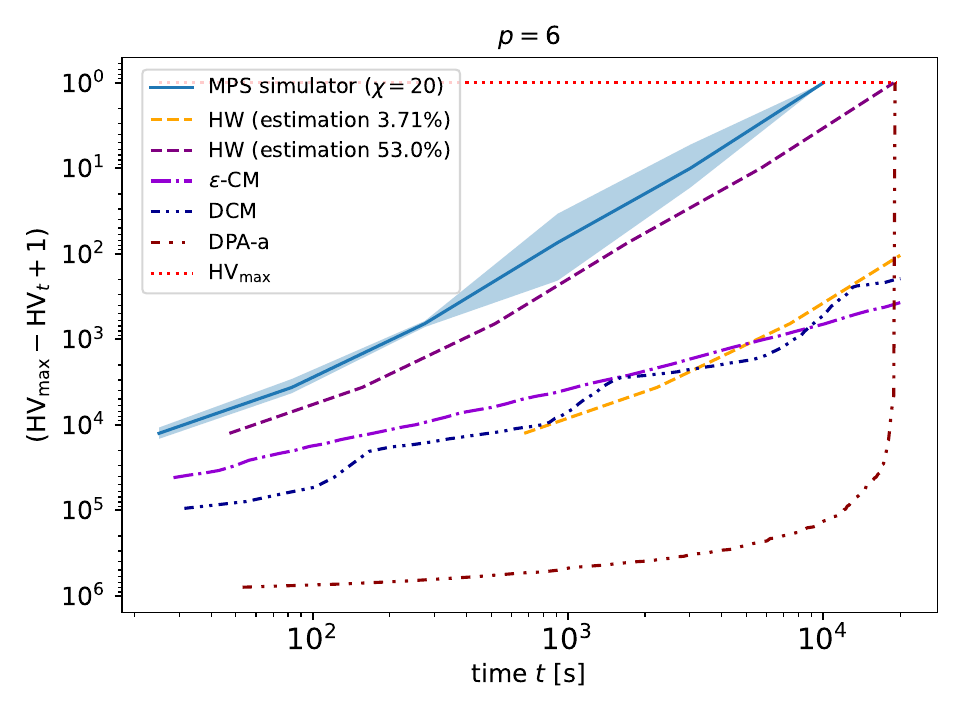}
    \caption{Comparison of classical and estimated quantum results for MO-MAXCUT with $m=4$ objective functions: Average progress of $\text{HV}_t$ of the MPS simulation ($\chi=20$, solid line), estimated hardware with 3.71\% fidelity, like \emph{ibm\_fez} (dashed orange line), estimated hardware with 53\% fidelity (dashed purple line), and the classical algorithms DCM, DPA-a, and $\epsilon$-CM (dotted-dashed lines). The shaded area denotes the minimum and maximum performance over the five repetitions of the MPS simulation. The best found solution $\text{HV}_{\text{max}}$ is indicated by the red dotted horizontal line.}
    \label{fig:42q_4o_hw_results}
\end{figure}

In addition to the classical algorithms, we run QAOA with $p=6$, $\num{20000}$ random $c$ vectors, and $\num{5000}$ shots per resulting circuit, i.e., we generate in total \num{100} million samples--four times more than for three objectives. Note that the circuit structure does not change compared to Sec.~\ref{sec:experiments_3o}. Assuming again \num{10000} shots per second this results in a total runtime of \num{10000}s. We run five repetitions of the MPS simulation with $\chi = 20$, which has been identified as sufficient in Sec.~\ref{sec:experiments_3o}. 
The resulting $\text{HV}_t$ for the different algorithms are shown in Fig.~\ref{fig:42q_4o_hw_results}.

In addition to the algorithms mentioned above, we estimate the performance of running on \emph{ibm\_fez} by scaling the time of the MPS simulation by the inverse of the estimated fidelity of 3.71\% (cf.~Tab.~\ref{tab:circuit_summary}), i.e., we multiply it by a factor of 26.954 and truncate after \num{20000}s. Following the previous discussion on the fidelity in Sec.~\ref{sec:experiments_3o}, this provides a lower bound on the performance to be expected when running the circuits on \emph{ibm\_fez}. It can be seen that after approximately \num{5000}s the predicted HV overtakes and then stays above the HV achieved by DCM and the $\epsilon$-CM. It is also above the HV achieved by DPA-a until the last few minutes of the runtime, when DPA-a accelerates towards $\text{HV}_{\text{max}}$.
Further, we find that with a hypothetical fidelity of 53\%, QAOA would achieve $\text{HV}_{\text{max}}$ in the same time as DPA-a, while dominating all algorithms throughout the whole runtime. 

We can see that the problem gets significantly more difficult for the classical algorithms when increasing the number of objective functions. The MPS simulation also finds the maximum HV, however, it clearly dominates all other algorithms over the whole runtime and achieves the maximum HV the fastest.

\section{Discussion and Conclusion}
\label{sec:conclusion}

We proposed and demonstrated an efficient quantum algorithm for Multi-Objective Optimization based on QAOA.
In particular, we tested our approach on both MPS simulations and real quantum hardware and showed that parameter transfer allows to scale QAOA for the considered MO-MAXCUT problem and that the algorithm has the potential to outperform classical benchmarks.
While the studied problem is still relatively simple, advanced classical algorithms already seem to struggle. As problem sizes, graph densities, and number of objective functions increase further, we expect the runtime of classical solvers to also significantly grow, implying a huge potential for the presented quantum algorithm as quantum computers continue to improve. 

There are several opportunities to further improve the proposed quantum algorithm and extend its applicability.
First, transfer of parameter and similar strategies to circumvent training need to be expanded to other problem classes, e.g., practically relevant problems in finance or logistics, where trade-offs between objective functions, such as risk and return, are omnipresent, or higher order binary optimization. Further directions to improve the algorithm include warm-starting, custom mixer designs, non-uniform sampling of weights or other scalarization strategies, as well as combining classical and quantum approaches. 
The whole approach leverages the assumption that QAOA returns a variety of good samples, which is closely related to the concept of fair sampling (e.g., how uniformly are different configurations that have the same objective value sampled). While we empirically confirm the generation of good and diverse samples, further theoretical understanding of QAOA and fair sampling is important, also for leveraging QAOA beyond MOO.
In addition to enhancing the quantum algorithm, there may also be a potential to improve the classical algorithms or benchmark additional ones on the considered problem, to further support the potential for quantum advantage.

Our algorithm also offers a new approach for constrained optimization. Adding constraints as objective functions and approximating the Pareto front allows to select those solutions that satisfy the original constraints. This can be used to study equality as well as inequality constraints. Particularly the latter are usually difficult to include in QUBOs without adding additional variables as they cannot be easily modeled as penalty terms. But also for equality constraints it might be beneficial to approach them from the MOO perspective since this eliminates the need to tune the involved penalty factors.

Last, we have shown that today's quantum computers allow forecasting the performance of (heuristic) algorithms on future devices with lower errors or even fault-tolerance. Similarly, we have shown the opposite direction, i.e., that simulations can be used to estimate the performance of algorithms on noisy quantum computers with different noise levels. This highlights that today's quantum computers are scientific tools to support algorithmic discovery and to augment theoretical research by empirical analyses.

\subsection*{Acknowledgment}
Ayse Kotil was employed at IBM Quantum and Zuse Institute Berlin at the time of the research and is now at PlanQC GmbH. 
The work of Ayse Kotil and Stephanie Riedmüller for this article at Zuse Institute Berlin in the Research Campus MODAL has been funded by the German Federal Ministry of Education and Research (BMBF) (Fund Nos.~05M14ZAM, 05M20ZBM).
E.P.~and S.E.~were supported by the U.S.~Department of Energy through the Los Alamos National Laboratory and the NNSA's Advanced Simulation and Computing Beyond Moore's Law Program at Los Alamos National Laboratory. Los Alamos National Laboratory is operated by Triad National Security, LLC, for the National Nuclear Security Administration of U.S.~Department of Energy (Contract No.~89233218CNA000001). LANL report number~LA-UR-25-22069. 
D.J.E.~acknowledges funding within the HPQC project by the Austrian Research Promotion Agency (FFG, project number 897481) supported by the European Union – NextGenerationEU.

\appendix

\section{Sampling of $c$ vectors}
\label{sec:simplex_uniform_sampling}

A $c$ vector, as introduced in Sec.~\ref{sec:classical_algorithms}, forms a convex combination of real numbers, i.e.,  $c \in [0, 1]^m$ with $\sum_i c_i = 1$. Only the non-negativity of vectors is strictly necessary. When sampling from an $m$-dimensional vector space $V\subset \mathbb{R}^m_{\geq 0}$ and projecting the samples onto the $(m-1)$-dimensional simplex by normalization, the vectors will not be distributed equally on the simplex. To obtain a balanced representation of different objectives, we follow the sampling strategy described in~\cite{Rubin1981}, which we hereby summarize. To obtain a single $c$ vector, $m-1$ numbers are uniformly sampled from $[0, 1]$ to form a set. Next, $0$ and $1$ are added to the set to extend it to a set of $m+1$ numbers. A second set is constructed by the differences of consecutive elements from the sorted set. This procedure guarantees that the second set consists of non-negative numbers summing up to $1$. Using this procedure will ensure a uniformly distributed set of $c$ vectors on the $(m-1)$-dimensional simplex.

\section{Classical Results for MO-MAXCUT}
\label{sec:classical_results}

In this section, we discuss several classical alternatives to the randomized $\epsilon$-CM for solving MO-MAXCUT with $m=3$ objective functions presented in Sec.~\ref{sec:experiments_3o}. It turns out that DPA-a, DCM, and the randomized $\epsilon$-CM performed best in our tests, which is the reason we use them as a classical reference in the main text.
The resulting HV for each algorithm is summarized in Tab.~\ref{tab:classical_results}, and the tested algorithms are discussed in the following.
We report the runtimes to generate the samples, without the post-processing to determine non-dominated points or corresponding HV. All runtimes are determined using an AMD EPYC 7773X CPU.

As a trivial baseline, we draw samples uniformly at random from $\{0, 1\}^{n}$ for $n = 42$. Since uniform sampling is cheap, we draw 25 million samples, like in the case of our quantum algorithm. We do not report a runtime, as the samples are generated almost immediately.

In addition, we test the randomized WSM. We use \num{400000} random $c$ vectors. It turns out that the WSM performs worse than, e.g., $\epsilon$-CM, which we explain by the presence of non-supported Pareto optimal points that the weighted sum approach cannot find.
This is supported by the fact that the WSM only returns 158 non-dominated points, i.e., over an order of magnitude less than the DPA-a and DCM.
Every MIP in the WSM takes on average 0.0039s to solve, which, for \num{400000} MIPs, results in a single core runtime of about 26 minutes.

Since every $c$ vector defines a new single objective MAXCUT problem, we also test the randomized WSM with the Goemans-Williamson (GW) algorithm \cite{goemans_williamson_1995_maxcut} as solver. For a given MAXCUT instance, the GW algorithm first maps it to a Semidefinite Program (SDP) that can be solved efficiently, and then applies randomized rounding to sample good solutions. The GW algorithm finds solutions that achieve on average at least 0.87856 of the MAXCUT value.
Here, we apply the same strategy as in our quantum algorithm: We sample \num{5000} random $c$ vectors and then draw \num{5000} random samples from the SDP solution, which leads to a total of 25 million samples.
We use CVXPY~\cite{cvxpy} to solve the resulting SDPs, which takes 0.6341s per SDP on average.
Thus, solving all SDPs on a single core takes about 53 minutes, where the runtime contribution of the randomized rounding is negligible.

The adapted version of the $\epsilon$-CM proposed in \cite{laumanns2006}, in the following referred to as ``Adapted $\epsilon$-CM'', generates suitable constraints during the process, so that the number of iterations is no input parameter. We tested it with a time limit of \num{10000}s.

As a representative of multi-objective integer optimization algorithms, we consider the openly accessible implementations of DPA-a \cite{Daechert2024} and DCM \cite{Boland2017}. Both DPA-a and DCM are exact algorithms that enumerate all non-dominated points by decomposing the objective space in search regions and solve the resulting subproblems by scalarization methods calling the solver CPLEX. Each found non-dominated point redefines the decomposition. For this kind of algorithm, there is a trade-off between the number and the complexity of the subproblems. Hence, their performance highly depends on the problem size and the number of objectives.

In DPA (Defining Point Algorithm), the search region is a union of non-disjoint rectangular sets. Those rectangular subregions are described by upper bound vectors based on the already generated non-dominated points. In each iteration, the IP restricted to one rectangular search region is solved by a scalarization method. We here choose the augmented (DPA-a) $\epsilon$-CM for solving each subproblem, i.e., $\epsilon$-CM with objective function
\begin{equation*}
    f_i(x) + \rho \sum_{j=1}^m f_j(x)
\end{equation*}
for a small augmentation parameter $\rho > 0$.
The algorithm guarantees that the number of local upper bound vectors is minimal and that each non-dominated point is generated exactly once.

In DCM (Disjunctive Constraint Method), the search region is decomposed using disjunctive constraints. Disjunctive constraints allow solving one IP call in several subregions by using additional binary variables. In a pure disjunctive constraint method, those constraints would decompose the search region in disjoint subregions. However, in DCM, the disjunctive constraints are only applied to the last generated non-dominated point. Thus, each subregion is defined by lower bounds, upper bounds, and a guiding point. The subproblems are solved by $\epsilon$-CM with objective function
\begin{equation*}
    \sum_{i=1}^m f_i(x).
\end{equation*}
DCM guarantees each non-dominated point to be generated exactly once, and it limits the number of subregions and the number of sets of disjunctive constraints.

Since the implementations of DPA-a and DCM do not allow objective functions taking continuous values, we scale the objective functions by a factor of 1k and 10k, and round them to gain integer values. This corresponds to representing the weights with a precision of $10^{-3}$ and $10^{-4}$. The precision strongly influences the performance. While a high precision can lead to very long runtimes, a low precision can cause a gap to the optimal HV.

\begin{table}
    \centering
    \begin{tabular}{lccc}
        Algorithm & HV & \#NDP & runtime [min] \\
        \hline
        DPA-a (10k)   & \num{43471.704} & 2067 & 3.6 \\
        DPA-a (1k)    & \num{43471.704} & 2063 & 3.6 \\
        DCM (1k)      & \num{43471.704} & 2063 & 8 \\
        $\epsilon$-CM & \num{43471.222} & 2054 & 167 \\
        GW algorithm  & \num{43443.382} & 1567 & 53 \\
        WSM           & \num{43088.010} & 158 & 26 \\
        Adapted $\epsilon$-CM & \num{39217.589} & 117 & time limit\\
        Random sampling    & \num{35870.921} & 220 & -- \\
        DCM (10k)   & \num{30445.587} & 4 & time limit \\
        
    \end{tabular}
    \caption{Results of classical algorithms: For every tested classical algorithm we report the achieved HV, the corresponding number of unique non-dominated points (\#NDP), and its runtime. The algorithms are sorted with respect to descending HV. We set a time limit of \num{10000}s and terminate the algorithms if they do not finish before. Since random sampling is very cheap to run, we do not provide a runtime.}
    \label{tab:classical_results}
\end{table}

\section{QAOA Parameter Optimization with JuliQAOA}
\label{sec:juli_qaoa}

The QAOA angle vectors $\vec{\beta}$, $\vec{\gamma}$ are optimized using \texttt{JuliQAOA} \cite{Golden_2023} on an ensembles of five different $27$-qubit problem instances each for the three- and four-objective case, see also Appendix~\ref{sec:transfer_of_parameters}.
\texttt{JuliQAOA} performs full statevector simulation, where the cost of all basis states for the optimization problem are precomputed and supplied to the simulation software (i.e., all $2^{n}$ variable configurations, with $n=27$). The optimization routine is performed using a single basin-hopping iteration per $p$, and \emph{angle extrapolation} is used to initialize the angle search at $p+1$ using the best angles found at the previous index $p$. Specifically, the best learned vector of angles found at $p$ is then directly substituted into the $p+1$ vector, and the two new parameters for the additional cost and mixer layer are randomly initialized. This entire vector, including the depth-$p$ angles, is then optimized. The Basin-hopping optimization \cite{Wales_1997} is used to search within the local vicinity of the extrapolated angles, see Ref.~\cite{Golden_2023} for more details. The choice of basin hopping as the optimization routine, and using a single iteration per $p$, was found to be the best performing set of parameter choices for general QAOA learning in Ref.~\cite{Golden_2023} (in particular, additional basin hopping iterations typically did not substantially improve the quality of the angles found), which is why we use these settings in this context. The single basin-hopping iteration refers to the basin-hopping implementation parameter where we only explore one local minima in the vicinity of the extrapolated angles. Similar ideas to QAOA angle extrapolation has been proposed in Ref.~\cite{Sack_2023} under terms such as stationary points of QAOA energy landscapes and transition states - the underlying idea of using local minima angles found at lower depth allowing for good heuristic angles being found efficiently at higher rounds. The search at $p=1$ is randomly initialized. \texttt{JuliQAOA} has been used for robust QAOA angle finding in several previous studies \cite{Pelofske_2024, 3SAT_high_round, Golden_2023_SAT}. Importantly, although \texttt{JuliQAOA} combines several high quality QAOA angle finding techniques, the resulting angles are not guaranteed to be globally optimal and therefore we consider these angles \emph{heuristic QAOA angles}. 

The minimum requirement we impose on the QAOA angle training is that each step of $p$ must reduce the expectation value, i.e., the energy of the cost Hamiltonian. Occasionally, for certain random initializations (of $p=1$), this requirement is not met during training -- if this occurs, then we simply terminate and re-attempt the simulation. 
To quantify QAOA performance we measure the (noiseless) expectation value of the optimized QAOA angles and then compute its approximation ratio, defined by,
\begin{equation}
\text{Approximation Ratio} = \frac{C_\text{Max} - C_e}{C_\text{Max} - C_\text{Min}}.
\label{equation:approximation_ratio}
\end{equation}
This requires finding the minimum ($C_\text{Min}$) and maximum ($C_\text{Max}$) cost value out of all $2^n$ possible configurations.
Here, $C_e$ denotes the cost of the energy. Importantly, this is an approximation ratio definition for a minimization problem, where lower cost denotes higher approximation ratio. The reason for this is because the Hamiltonian cost function that QAOA is trained on is an energy minimization problem - but this is directly equivalent to the maximization problem of (weighted) MAXCUT. Note that this is an unconventional approximation ratio definition so as to accommodate negative, and positive, energies of the MAXCUT Hamiltonian. This definition also means that random sampling on average will result in an approximation ratio of $\approx 0.5$. 

\begin{figure}[ht!]
    \centering
    \includegraphics[width=1.0\linewidth]{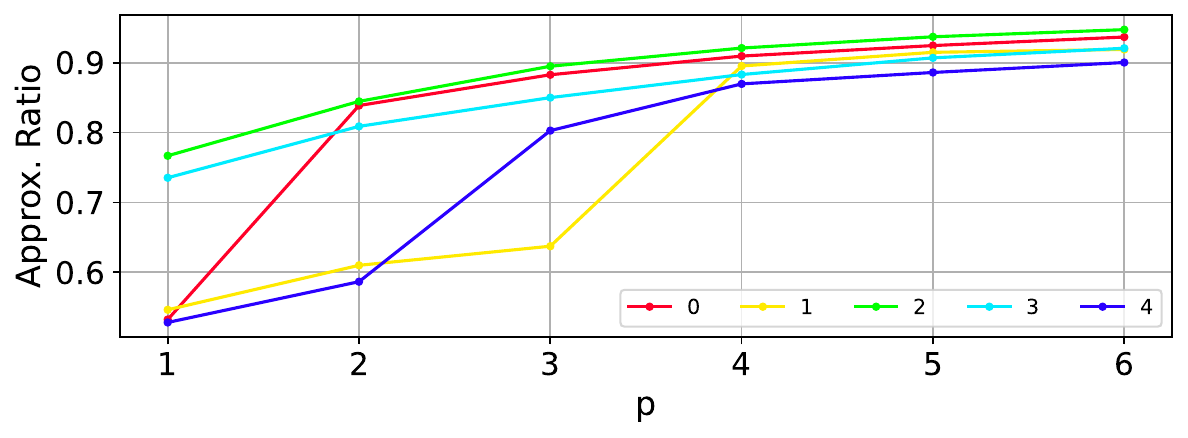}
    \includegraphics[width=1.0\linewidth]{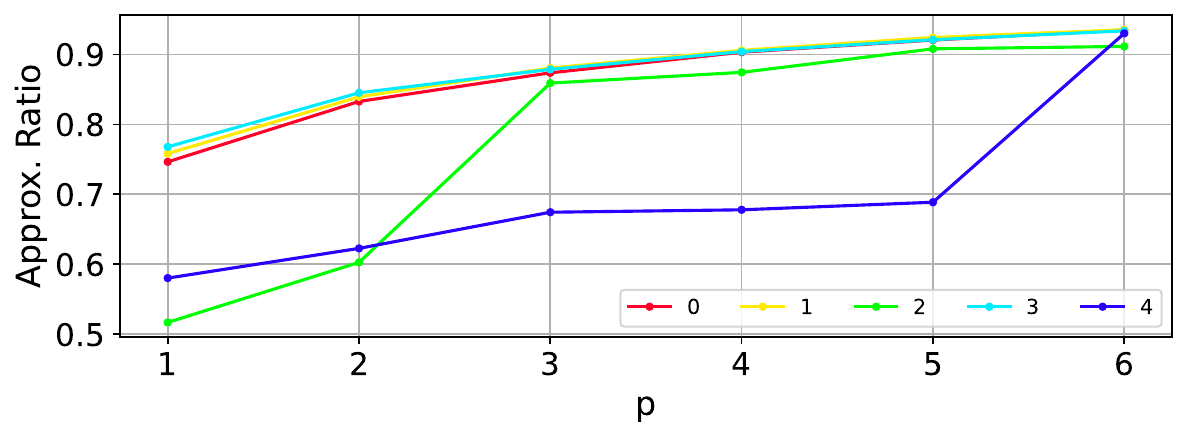}
    \caption{Approximation ratio of the learned QAOA angles on the set of $5$ random weighted MAXCUT instances with combined $3$-objectives (top) and $4$-objectives (bottom). The y-axis plots the noiseless approximation ratio and the x-axis is the QAOA depth $p$. Each line shows the approximation ratio for one of the five different distinct problem Hamiltonians, the (arbitrary) index for the problem Hamiltonian is shown by the legend(s). \texttt{JuliQAOA}'s angle learning is able to provide a consistent improvement of solution quality; at $p=6$, QAOA achieves an approximation ratio better than $0.9$. }
    \label{fig:QAOA_approximation_ratio}
\end{figure}

Fig.~\ref{fig:QAOA_approximation_ratio} shows the learned QAOA approximation ratios for the five different random problem Hamiltonians with $3$-objectives and five different random Hamiltonians with $4$-objectives. 
For our experiments we selected the parameters resulting from the first random instance.

\section{Transfer of QAOA Parameters and Compiled QAOA Circuit Structure\label{sec:transfer_of_parameters}}

In our framework, problem Hamiltonians are constructed via the weighted sum of the problem graphs. The weighted sum does not change the structure of the problem Hamiltonians, i.e., the interacting qubit pairs remain the same. Only the coefficients of the individual Pauli terms differ with varying weight sets. 
Since the graph structure does not change and the weights are randomly selected, intuition suggests that we can transfer the parameters between different graph instances. 
We show this empirically as we do not have a formal proof.
We examine MAXCUT instances on 27-node hardware-native graphs (i) with five different realizations with edge weights sampled from $\mathcal{N}(0, 1/3)$ for the case of three objective functions, and (ii) with five instances with weights sampled from $\mathcal{N}(0, 1/4)$ for the case of four objective functions.

\begin{figure*}[ht!]
    \centering
    \includegraphics[width=0.32\linewidth]{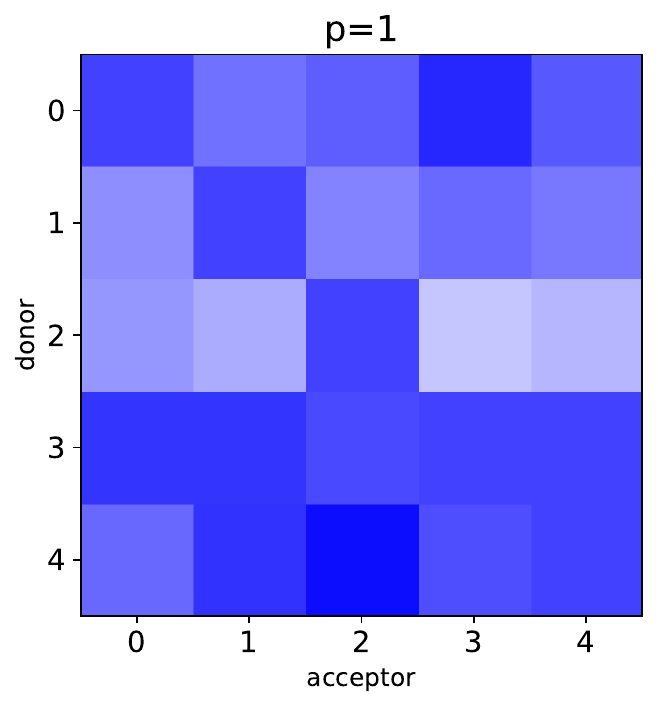}
    \includegraphics[width=0.32\linewidth]{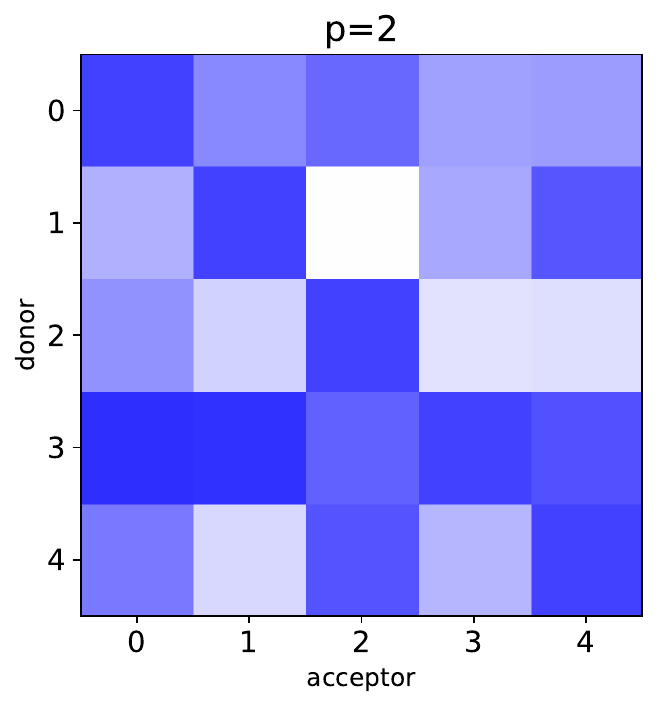}
    \includegraphics[width=0.32\linewidth]{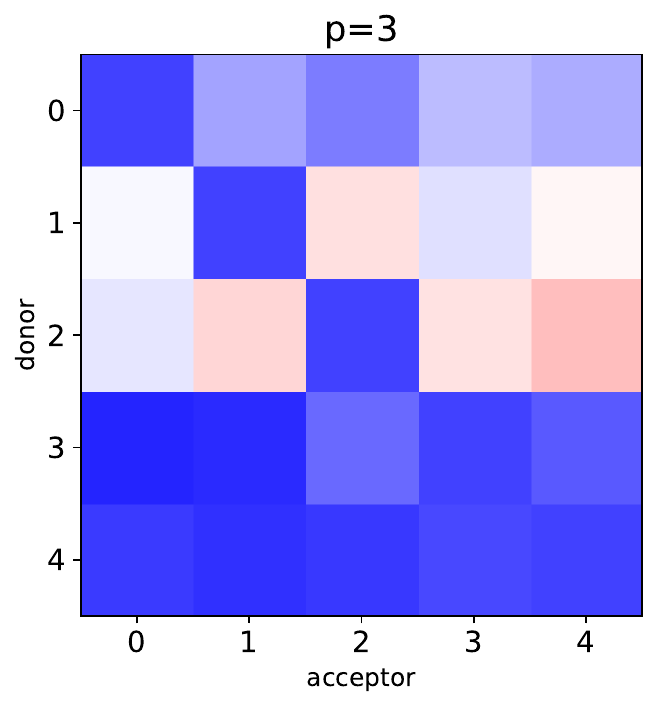}
    \includegraphics[width=0.32\linewidth]{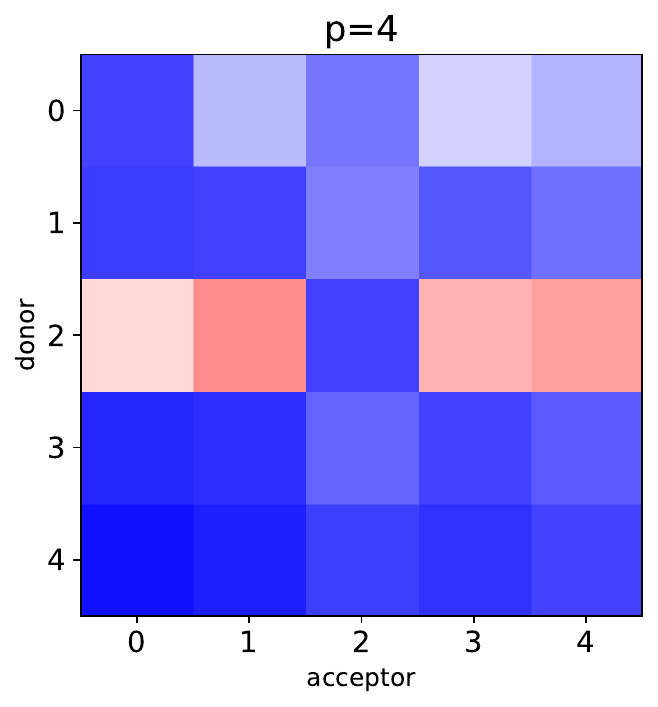}
    \includegraphics[width=0.32\linewidth]{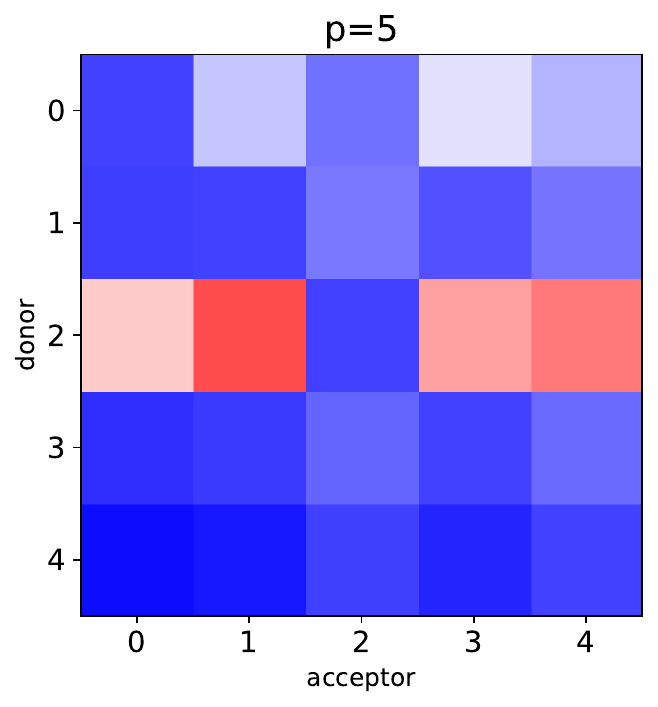}
    \includegraphics[width=0.32\linewidth]{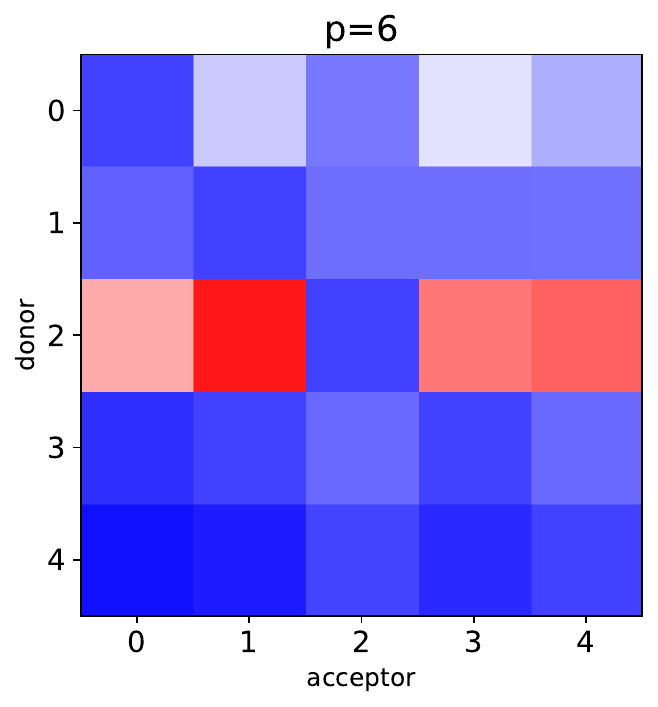}
    \includegraphics[width=0.35\linewidth]{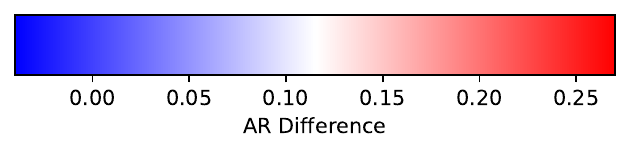}
    \caption{QAOA parameter transfer matrices for $p=1$ (top left) to $p=6$ (bottom right). For a set of $5$ fixed problem instances, $5$ random coefficient choices each for hardware-native problem instances with $3$ combined objective functions, \texttt{JuliQAOA} is used to optimize the angles for each problem instance. Each matrix plots the difference between the approximation ratio of QAOA using the angles trained for that problem instance and the approximation ratio of QAOA using those same angles applied to a different (unseen during training) problem instance; each unique instance is named on y-axis and the trained parameters for that instance are then applied to the other $4$ problem instances (x-axis). Positive values mean that the original approximation ratio was better than the approximation ratio on the new (unseen) problem. Negative value means the parameters worked even better on the new (unseen) problem. Diagonal entries, by definition, have an approximation ratio difference of zero. The absolute possible range of this approximation ratio difference is $[-1, 1]$. 
    }
    \label{fig:parameter_concentration}
\end{figure*}

\begin{figure*}[ht!]
    \centering
    \includegraphics[width=0.32\linewidth]{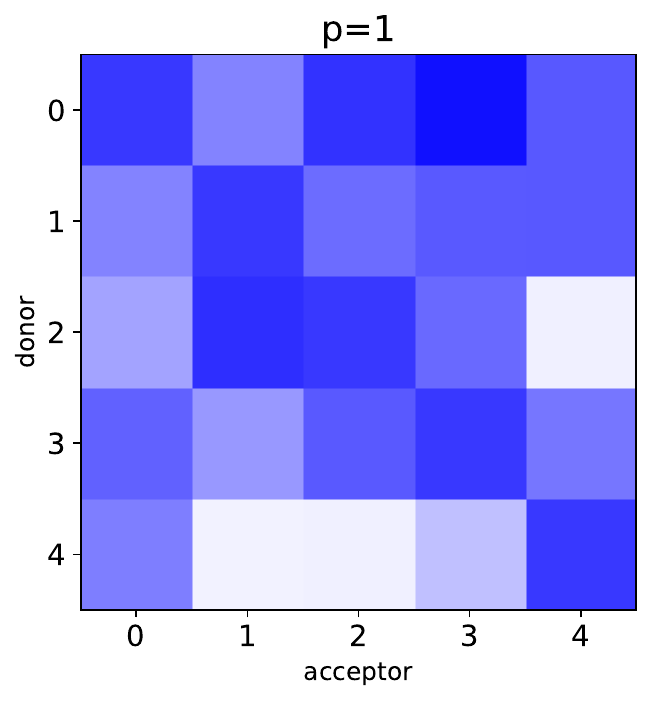}
    \includegraphics[width=0.32\linewidth]{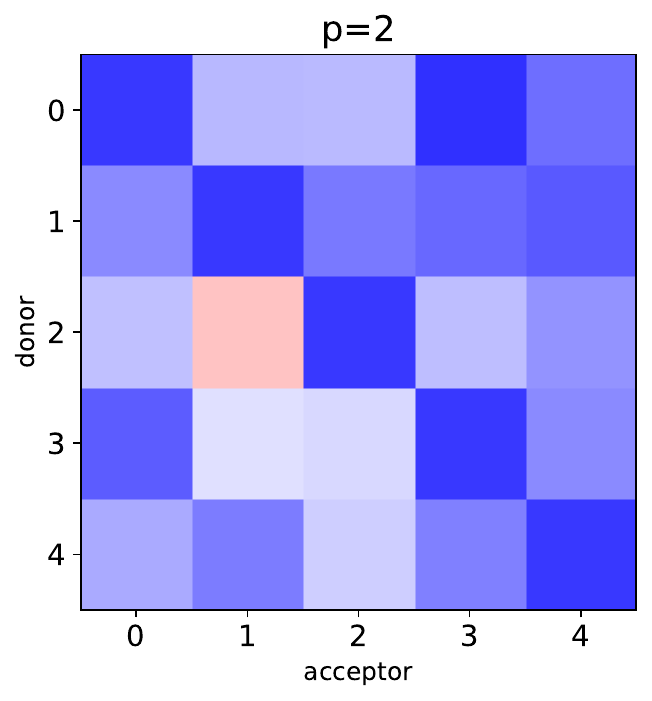}
    \includegraphics[width=0.32\linewidth]{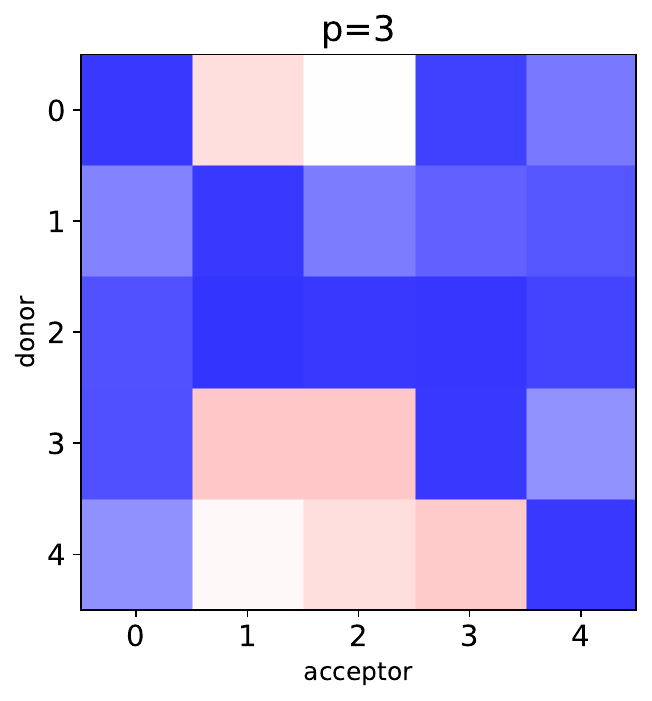}
    \includegraphics[width=0.32\linewidth]{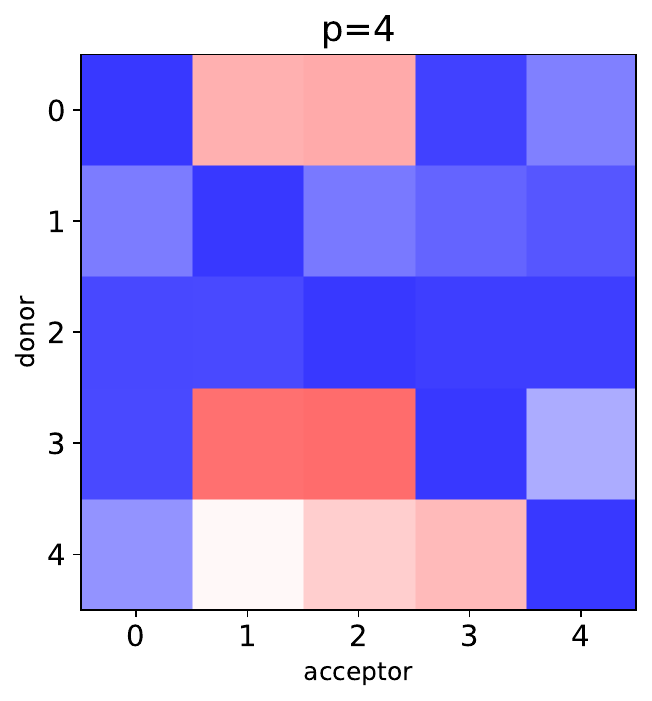}
    \includegraphics[width=0.32\linewidth]{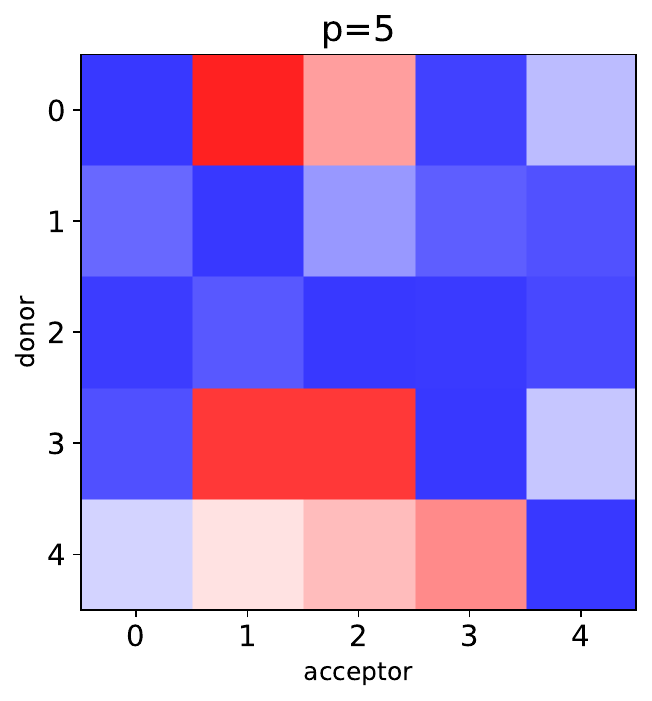}
    \includegraphics[width=0.32\linewidth]{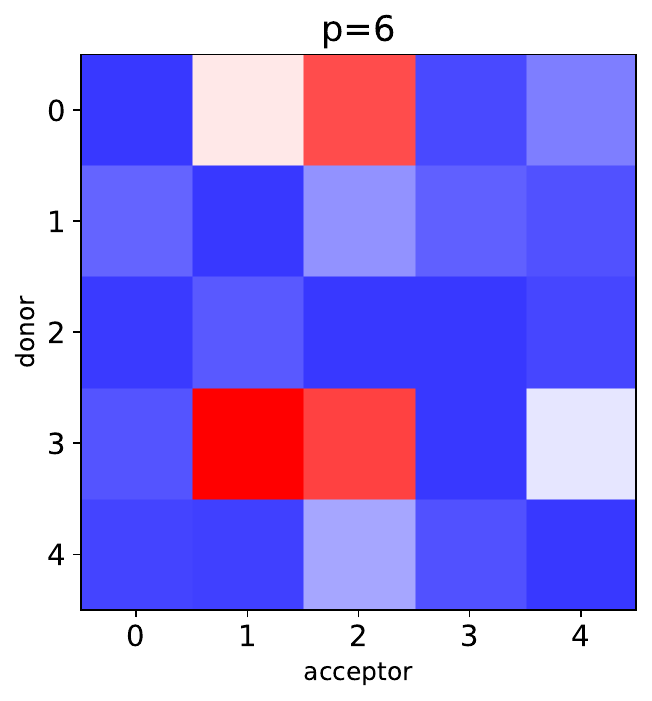}
    \includegraphics[width=0.35\linewidth]{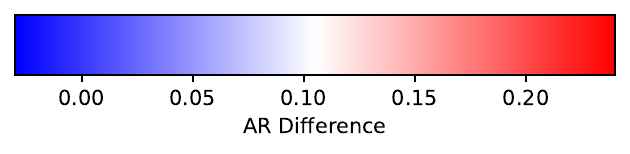}
    \caption{QAOA parameter transfer matrices for $p=1$ (top left) to $p=6$ (bottom right) on five instances of $4$-objective weighted MAXCUT heavy-hex instances. Same format as Figure~\ref{fig:parameter_concentration}, but now working on $4$-objective instances as opposed to $3$. Negative approximation ratio difference denotes that the trained parameters on the donor instance worked even better on the acceptor angles (this, as expected, happens rarely). Positive values mean the approximation ratio of the trained angles on the donor instance are better than the approximation ratio on the acceptor instance - the more positive the values are, the worse the parameter transfer worked.}
    \label{fig:parameter_concentration_4_objectives}
\end{figure*}

Fig.~\ref{fig:parameter_concentration} and \ref{fig:parameter_concentration_4_objectives} show parameter transfer of optimized QAOA angles from $27$-qubit instances, as described in Sec.~\ref{sec:juli_qaoa}, onto $27$-qubit instances, for 3 and 4 objectives, respectively. 
Here we are showing the approximation ratio difference between the noiseless expectation value of original (donor) problem and the new problem (acceptor). 
As before, these QAOA simulations use the single-objective weighted MAXCUT problem Hamiltonian formed from linear combinations of either $3$ or $4$ other weighted MAXCUT Hamiltonians.
This means that the energy of the classical Hamiltonian that the noiseless QAOA numerical simulation computes is measured on this linear combination of weighted MAXCUT Hamiltonians. Overall these small scale numerical experiments demonstrate that QAOA parameter re-use is reasonable to utilize across this set of problem instances -- with the exception of one particular instance which did not transfer well (middle instance in the matrices of Fig.~\ref{fig:parameter_concentration}). In practice, we use the learned QAOA angles from the first random problem instance (donor 0) when executing the QAOA circuits on the noisy quantum hardware. 

\section{Mixed Integer Programming Formulation for Weighted-MAXCUT}
\label{sec:mip_maxcut}

In this section we briefly introduce the basic MIP formulation for Weighted-MAXCUT that we use in the $\epsilon$-CM. Suppose a weighted graph $(\mathcal{V}, \mathcal{E}, w)$, then MAXCUT can be modeled as \cite{Charfreitag2023}
\begin{align*}
    \max_{x, s}\;& \sum_{(k,l) \in \mathcal{E}} w_{ij} e_{ij} \\
    \text{subject to:\;}
    & x_{k} \in \{0, 1\},\; &\forall k \in \mathcal{V},\\
    & e_{kl} \in \{0, 1\},\, &\forall (k, l) \in \mathcal{E},\\
    & e_{kl} \leq x_k + x_l,\, &\forall (k, l) \in \mathcal{E},\\
    & e_{kl} \leq 2 - x_k - x_l,\, &\forall (k, l) \in \mathcal{E},\\
    & e_{kl} \geq x_k - x_l,\, &\forall (k, l) \in \mathcal{E},\\
    & e_{kl} \geq x_l - x_k,\, &\forall (k, l) \in \mathcal{E}.\\
\end{align*}
The node variables $x_k$ indicate whether node $k \in \mathcal{V}$ is in the first or second class. The edge variables $e_{kl}$ for $(k, l) \in \mathcal{E}$ determine whether the corresponding edge is cut or not.
The last two constraints are not needed in case of positive weights.
However, if negative weights can occur, the lower bounds need to be added to enforce that $e_{kl} = 1$ is equivalent to $x_k \neq x_l$ for all $(k, l) \in \mathcal{E}$.
Further, suppose we have multiple sets of weights $w_{kl}^i$, for $i = 1,\ldots, m$. Then we can add additional constraints $\sum_{(k,l)} w_{kl}^i e_{kl} \geq \epsilon_i$ to enforce a lower bound on the individual MAXCUT objectives and adjust the weights in the objective to represent convex combinations of the $w_{kl}^i$.

\section{Convergence \& Runtime Analysis of MPS Simulations}
\label{sec:mps_convergence}

\begin{figure}
    \centering
    \includegraphics[width=1\linewidth]{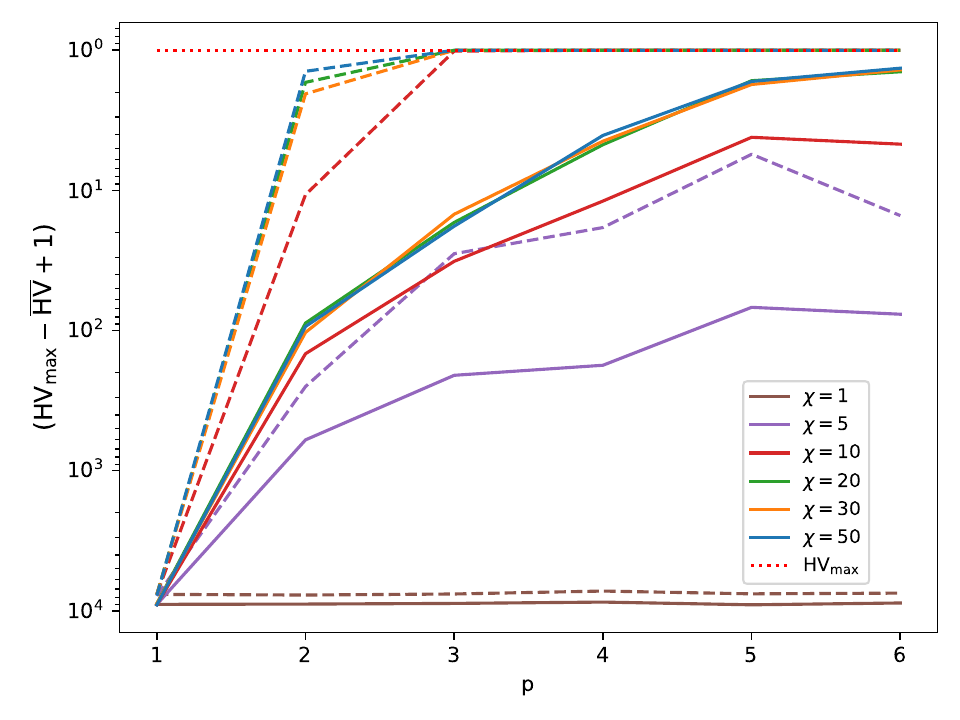}
    \caption{Convergence of MPS simulations: This plot shows the different HV time-averages $\overline{\text{HV}}$ as well as maximum achieved HV for different bond dimensions $\chi$ and QAOA repetitions $p$. As before, we show $\text{HV}_{\text{max}} - \overline{\text{HV}} + 1$ with a logarithmic and reversed y-axis. All simulations were run with \num{5000} random $c$ vectors and \num{5000} shots for each resulting circuit.}
    \label{fig:42q_mps_sim_results_avg}
\end{figure}

In this section, we analyze the convergence of the MPS simulations with respect to the bond dimension. To this extent, we consider the 42-node 3-objective MO-MAXCUT problem instance and evaluate the average HV over time, i.e., $\overline{\text{HV}} = 1/T \sum_{t=1}^T \text{HV}_t$, for different bond dimensions $\chi$ and QAOA repetitions $p$, as well as the resulting maximum HV for each setting. 
The HV does not significantly change anymore for $\chi \geq 20$, see Fig.~\ref{fig:42q_mps_sim_results_avg}, i.e., we assume this is sufficient to represent the considered 42-qubit quantum circuits.

In addition, we analyze the runtime of the MPS simulations to generate the 25 million samples from the 42-qubit QAOA circuits for different bond dimensions $\chi$ and different number of repetitions $p$, as discussed in Sec.~\ref{sec:experiments}. We measure the single-core runtime on an \emph{Apple M1 Max} CPU with 32 GB. This can be easily translated to multi-core runtime assuming no memory bottlenecks. Tab.~\ref{tab:mps_runtimes} shows the resulting runtimes.
It can be seen that the runtimes quickly exceed multiple hours for larger $\chi$ and $p$. If we consider $p \geq 3$ with $\chi = 20$, i.e., the settings that first start to achieve $\text{HV}_{\text{max}}$, the single-core runtimes are around three hours, i.e., more than 4-times slower than generating the samples using the real quantum computer.
While the simulation can be easily parallelized, the results also very clearly show the exponential growth of the runtimes with increasing bond dimension, which implies that this will quickly become intractable for larger circuits.

\begin{table}
    \centering
    \begin{tabular}{cc|cccccc}
                     &   &  \multicolumn{6}{c}{$\chi$}    \\
                     &   &  1 &  5 & 10 &  20 &  30 &  50 \\
                     \hline
\multirow{6}{*}{$p$} & 1 &  9 & 55 & 76 & 106 & 134 & 133 \\
                     & 2 &  9 & 60 & 84 & 168 & 309 & 679 \\
                     & 3 &  9 & 59 & 86 & 178 & 341 & 819 \\
                     & 4 &  9 & 59 & 87 & 191 & 355 & 866 \\
                     & 5 & 10 & 59 & 87 & 186 & 364 & 910 \\
                     & 6 &  9 & 59 & 87 & 190 & 370 & 962 \\
    \end{tabular}
    \caption{Runtime of MPS simulations: Runtime in minutes to generate 25 million samples for different bond dimensions $\chi$ and QAOA repetitions $p$ using a single core on an \emph{Apple M1 Max} CPU.}
    \label{tab:mps_runtimes}
\end{table}

\bibliography{references}

\end{document}